\documentclass[fleqn,usenatbib]{mnras}
\usepackage{amsmath}
\usepackage{newtxtext,newtxmath}

\usepackage[T1]{fontenc}

\DeclareRobustCommand{\VAN}[3]{#2}
\let\VANthebibliography\thebibliography
\def\thebibliography{\DeclareRobustCommand{\VAN}[3]{##3}\VANthebibliography}

\usepackage{graphicx}	
\usepackage{amsmath}	

\usepackage{amssymb}	
\usepackage[dvipsnames]{xcolor}
\usepackage[utf8]{inputenc}
\usepackage{multirow}
\usepackage{orcidlink}

\title[Influence of SEDs on the Ly$\alpha$ forest]{Influence of the Spectral Energy Distribution of Reionization-Era Sources on the Lyman-$\alpha$ Forest}

\author[Basu et al.]{
Arghyadeep Basu \orcidlink{0000-0001-8104-9751} $^{1,2,3}$\thanks{E-mail: basu@mpa-garching.mpg.de}\thanks{E-mail: basu.arghyadeep@yahoo.in},
Benedetta Ciardi \orcidlink{0000-0002-5037-310X}$^{1}$,
James S. Bolton \orcidlink{0000-0003-2764-8248}$^{4}$,
Matteo Viel \orcidlink{0000-0002-2642-5707}$^{5,6,7,8,9}$,
Enrico Garaldi \orcidlink{0000-0002-6021-7020}$^{10,11}$
\\
$^{1}$Max-planck-Institut f$\ddot{u}$r Astrophysik, Karl-Schwarzschild-Strasse 1, D-85741, Garching, Germany\\
$^{2}$Ludwig-Maximilians-Universität München (LMU), Geschwister-Scholl-Platz 1, 80539 München, Germany\\
$^{3}$Univ Lyon, Univ Lyon1, Ens de Lyon, CNRS, CRAL UMR5574, F-69230, Saint-Genis-Laval, France\\
$^{4}$School of Physics and Astronomy, University of Nottingham,
University Park, Nottingham, NG7 2RD, UK\\
$^{5}$SISSA - International School for Advanced Studies, Via Bonomea 265, 34136 Trieste, Italy\\
$^{6}$INFN – National Institute for Nuclear Physics,
Via Valerio 2, I-34127 Trieste, Italy\\
$^{7}$IFPU, Institute for Fundamental Physics of the Universe,
via Beirut 2, 34151 Trieste, Italy\\
$^{8}$INAF, Osservatorio Astronomico di Trieste,
Via G. B. Tiepolo 11, I-34131 Trieste, Italy\\
$^{9}${ICSC - Centro Nazionale di Ricerca in High Performance Computing,
Big Data e Quantum Computing, Via Magnanelli 2, Bologna, Italy}\\
$^{10}$Kavli IPMU (WPI), UTIAS, The University of Tokyo, Kashiwa, Chiba 277-8583, Japan\\
$^{11}$Center for Data-Driven Discovery, Kavli IPMU (WPI), UTIAS, The University of Tokyo, Kashiwa, Chiba 277-8583, Japan\\
}

\date{Accepted 2026 January 22. Received 2026 January 20; in original form 2025 September 18.}

\pubyear{2015}

\begin{document}
\label{firstpage}
\pagerange{\pageref{firstpage}--\pageref{lastpage}}
\maketitle

\begin{abstract} 
Interpreting Lyman-$\alpha$ forest properties during the epoch of reionization requires assumptions about the spectral energy distribution (SED) of ionizing sources. These are often simplified to blackbody or power-law spectra, potentially overlooking contributions from high-energy processes. In this work, we investigate how different SED models of reionization-era sources shape the thermal and ionization state of the intergalactic medium (IGM) and imprint on the Ly$\alpha$ forest during the late stages of reionization. We perform 3D radiative transfer simulations with \texttt{CRASH}, post-processed on \texttt{Sherwood}-type hydrodynamical outputs, exploring both physically motivated SEDs ie. SED including X-ray binaries, Bremsstrahlung from shock-heated interstellar medium, and binary stars, and idealized blackbody and power-law spectra. While the large-scale morphology of ionized regions is broadly similar across all models, harder spectral components extend partially ionized zones, produce larger He~{\sc iii} regions, and heat the surrounding IGM. By adopting simplified spectra there is the risk of underestimating the contribution of high-energy sources, which for most models we consider are found to alter the effective optical depth, the flux power, and the local transmissivity within the current $\sim 1 \sigma$ measurement uncertainties.
The differences across models are most pronounced in the behavior of the proximity zone and in the power at intermediate scales, offering the most promising diagnostics to disentangle source populations. With upcoming high-precision measurements from \texttt{ELT} and \texttt{DESI}, realistic SED modelling will be essential for robustly connecting Ly$\alpha$ forest observations to the sources driving the end of reionization.

\end{abstract}




\section{Introduction}
\label{section:1}
The growth of primordial density fluctuations from gravitational instabilities gave rise to the formation of the first stars and galaxies, which started to emit  photons with enough energy to ionize the surrounding intergalactic medium (IGM) and initiate the last global phase transition of the universe, namely the Epoch of Reionization (EoR; \citealt{loeb2001}). Observations of high-$z$ QSO absorption spectra \citep{gunn1965,Fan2006,McGreer2015,bosman2018,Davies2018,Bosman2022}, Lyman-$\alpha$ emitters (LAEs; \citealt{Pentericci2014,Mason2018,jung2022,tang2024,nakane2024,napolitano2024,saxena2024,chen2024,ning2024}), information of the IGM thermal properties from QSO spectra \citep{Bolton2012,Raskutti2012,Gaikwad2020}, and of the Thompson scattering optical depth from Cosmic Microwave Background (CMB;  \citealt{Komatsu2011,planck2016,pagano2020}) 
indicate that reionization is an extended process, largely complete by $z\approx 5.5$ \citep{kulkarni2019,nasir2020}. 


This abundance of observations is accompanied by an increasingly sophisticated modeling of the reionization process, employing a variety of approaches, ranging from semi-analytic and semi-numeric models \citep{zhou2013,mesinger2007,santos2010,hutter2021,choudhury2022}, to full radiation hydrodynamic simulations \citep{Baek2010, gnedin2014,finlator2018,rosdahl2018,ockvirk2020,trebitsch2021,kannan2022,garaldi2022,bhagwat2024}.
Although in most literature galaxies are thought to be the main drivers of reionization  \citep[e.g.][]{BeckerBolton2013,Bouwens2015,Robertson2015,Madau2017,dayal2020,hassan2018,Marius2020,eide2020b,kannan2022}, it is still unclear which are the properties of those contributing the most (e.g. small vs. massive galaxies; \citealt{Naidu2020,Naidu2021,Matthee2022}). Additionally, the recent detection of a larger than expected number density of faint quasars at intermediate (i.e. $z\approx4$; \citealt{Giallongo2015,Giallongo2019,Boutsia2018}) and high (i.e. $z>6$;  \citealt{Weigel2015,McGreer2018,Parsha2018,Akiyama2018,Harikane2023,Maiolino2023,Goulding2023,Larson2023,Juodzbalis2023,Greene2023}) redshift, have renewed the interest in quasars as possible strong contributors to the ionization budget (\citealt{grazian2024,madau2024}, but see also \citealt{Asthana2024}).

Whatever the main source of ionizing photons is, the physical properties of the IGM in terms of ionization and thermal state are influenced by a variety of sources, including the above mentioned stars and quasars, but also X-ray binaries (XRBs; \citealt{mirabel2011,fragos2013,fragosa2013,madaufragos2017,sazonov2017}), Bremsstrahlung emission from shock-heated interstellar medium (ISM; \citealt{Chevalier1985,suchkov1994,strickland2000}), low energy cosmic rays \citep{nath1993,sazonov2015,leite2017,owen2019,bera2023,yokoyama2023,gessey2023,carvalho2024}, self-annihilation or decay of dark matter \citep{liu2016,cang2023} and plasma beam instabilities in TeV blazars \citep{chang2012,puchwein2012,lamberts2022}. 

As discussed for example in \citet{Marius2018} and \citet{Marius2020} (but see also, \citealt{Baek2010,Venkatesan2011,Grissom2014,enrico2019}), the more energetic sources are responsible for the partially ionized and warm gas found outside of the regions fully ionized by stars. Additionally, they can also increase the IGM temperature in comparison to the one obtained in star driven reionization scenarios. 
Their impact on studies of the 21cm signal from neutral hydrogen at high-$z$ has been widely recognized and investigated \citep{ross2017,qingbo2020,qingbo2021,qingbo2022,kamran2022,noble2024}.
Such high-energy sources may also impact the Lyman-$\alpha$ (Ly$\alpha$) forest by broadening thermal widths and lowering the IGM neutral fraction, thereby increasing the transmission flux.

Here, in this paper, we present a systematic study of the impact of different  source spectral energy distributions by post-processing \texttt{Sherwood}-type cosmological hydrodynamic simulation \citep{Bolton2017} using the 3D multi-frequency radiative transfer code CRASH \citep[e.g.][]{ciardi2001,maselli2003,maselli2009,graziani2013,hariharan2017,glatzle2019,glatzle2022} which follows the formation and evolution of ionized hydrogen and helium along with the photoheating of IGM utilising different assumptions of the properties of ionizing sources. The paper is organized as follows: Section~\ref{methodology} outlines the simulations, Section~\ref{results} presents our findings, and Section~\ref{discussion} concludes with a broader interpretation of the results and a summary.

\section{Methodology}
\label{methodology}
Here, we describe how we have post-processed the outputs of a hydrodynamical simulation (Section \ref{hydrosim}) with the multi-frequency radiative transfer (RT) code \texttt{CRASH} (Section \ref{crash}) to obtain a suite of simulations of reionization (Section \ref{section:2.3}).

\subsection{The cosmological hydrodynamic simulation}
\label{hydrosim}

To obtain the gas and source distribution and properties, we have used outputs at redshift 5.16 to 15 from \texttt{Sherwood}-type cosmological hydrodynamic simulation \citep{Bolton2017} performed in a comoving cubic box of size 20 $h^{-1}$ cMpc \footnote{The simulation volume here represents a compromise between capturing large-scale structure and achieving the spatial resolution required for full 3D radiative transfer. While such a volume does not fully sample the rarest, large-scale Ly$\alpha$ opacity fluctuations, the analysis in this work focuses on statistics that are expected to be robust to finite-volume effects. Moreover, the primary aim of this study is to quantify relative differences between models with different ionizing SEDs using the same underlying density field and radiative transfer framework, rather than to fully reproduce the observed Ly$\alpha$ opacity distribution. For this purpose, a 20~$h^{-1}$~cMpc volume is sufficient to robustly capture the bulk Ly$\alpha$ forest statistics considered here.}
with the parallel Tree-PM smoothed particle hydrodynamics (SPH) code \texttt{p-GADGET-3},   which is an updated version of \texttt{GADGET-2} \citep{Volker2001,gadget2}. The box contains $2\times1024^{3}$ gas and dark matter particles, each of mass $m\rm{_{gas}}=9.97 \times 10^{6}\, h^{-1} M_{\odot}$ and  $m\rm{_{DM}}=5.37 \times 10^{7}\, h^{-1} M_{\odot}$, respectively. Haloes are identified using the friends-of-friends algorithm with a linking length of $0.2$ times the mean inter-particle separation. The gravitational softening length for the simulation is set to 1/30th of the mean linear interparticle spacing, and star formation is incorporated using a simple and computationally efficient prescription which converts all gas particles with overdensity $\Delta > 10^{3}$ and temperature $T<10^{5} \rm{K}$ into collisionless stars \citep{Viel2004,Viel2013}. 
Note that, due to this simplified treatment, the simulation does not self-consistently model star formation and stellar feedback. However, the Ly$\alpha$ forest statistics considered here primarily probe the low-density diffuse IGM ($\Delta \lesssim 10$), where gas dynamics are dominated by gravitational collapse and photoheating, and are only weakly sensitive to the detailed modelling of star formation and feedback. Feedback effects are largely confined to high-density regions near galaxies, while the forest absorption at $z\gtrsim4$ arises predominantly from gas well beyond the virial radii of typical sources \citep{Viel2013,Keating2016,Nasir2017}.
The simulations adopt a spatially uniform, optically thin ionizing background based on the \citet{Haardt2012} model, which is turned on at $z=9$ and evolves with redshift to set the baseline ionization and thermal state of the diffuse IGM, which is used to set the gas pressure smoothing scale in the simulation. Note that this is used to approximately capture the impact on gas dynamics during the simulation, but is then discarded and recomputed more accurately in this work, as discussed in the following sections.
 The model adopts a Planck 2013 consistent cosmology  \citep{planck2013} with $\rm{\Omega_{m}=0.308}$, $\rm{\Omega_{\Lambda}=0.692}$, $\rm{\Omega_{b}=0.0482}$, $\rm{\sigma_{8}=0.829}$, $\rm{\mathit{n}_{s}=0.963}$, $h=0.678$, where the symbols have their usual meaning.
Throughout the paper, the gas is assumed to be of primordial composition with a hydrogen and helium number fraction of 0.92 and 0.08, respectively. 

\begin{figure} 
\centering
    \includegraphics[width=\columnwidth]{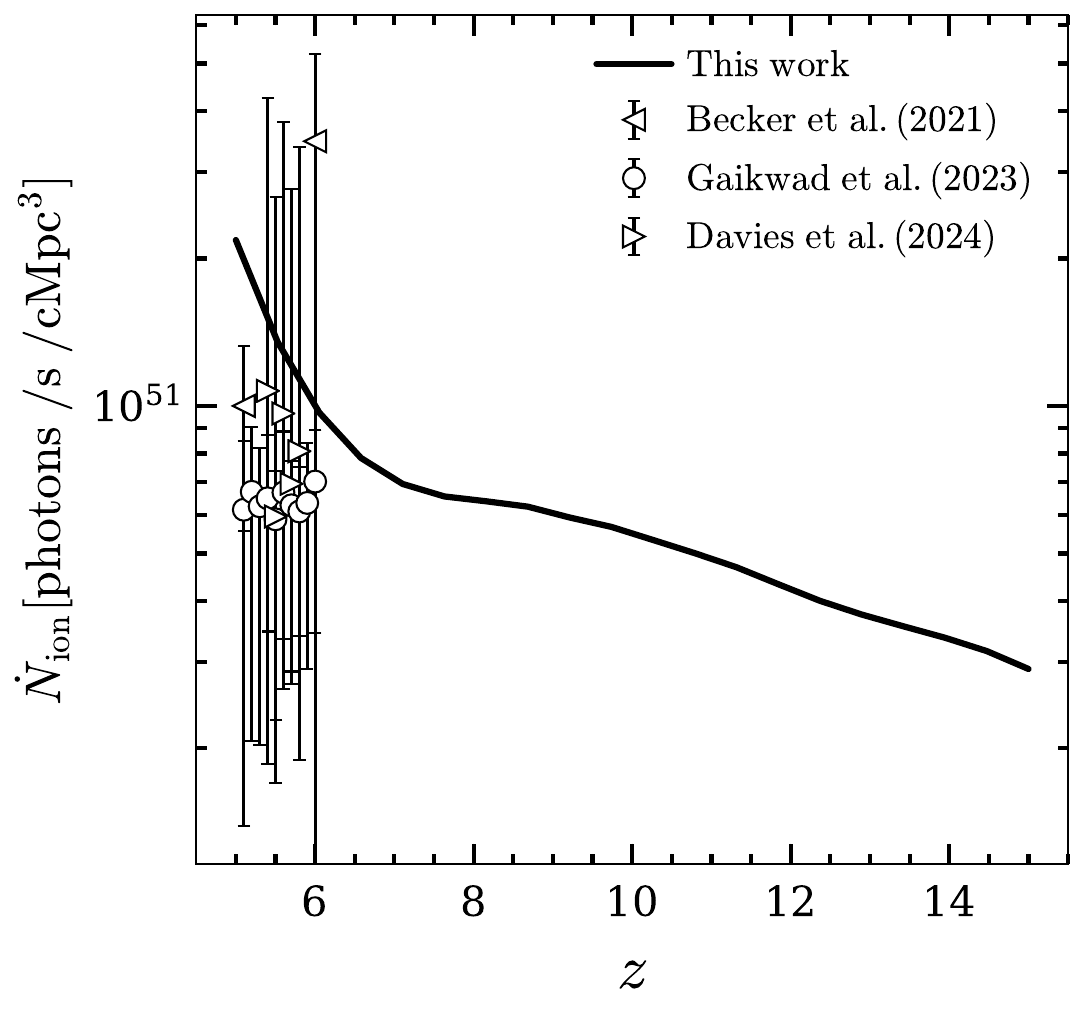}  
    \caption{Redshift evolution of the total ionizing photon production rate adopted in the study (black solid line) with a compilation of observational constraints \citep{becker2021,gaikwad2023,davies2024}.}
    \label{fig:siminput_emiss}
\end{figure}

\subsection{The radiative transfer code \texttt{CRASH}}
\label{crash}
The radiative transfer of ionizing photons through the IGM is implemented by post-processing the outputs of the hydrodynamic simulation using the 3D non-equilibrium and multi-frequency code \texttt{CRASH}  \citep[e.g.][]{ciardi2001,maselli2003,maselli2009,graziani2013,hariharan2017,glatzle2019,glatzle2022}. The code evaluates self-consistently the ionization states of hydrogen and helium, as well as the evolution of gas temperature. \texttt{CRASH} employs a Monte Carlo-based ray tracing approach, where the propagation of ionizing radiation is represented by multi-frequency photon packets that dynamically trace the spatial and temporal distribution of radiation within the simulation volume. The version of \texttt{CRASH} employed here has a comprehensive treatment of UV and soft X-ray photons, including X-ray ionization, heating, and secondary electron interactions \citep{graziani2018}. For an in-depth understanding of \texttt{CRASH}, we encourage readers to refer to the foundational papers.

\subsection{The radiative transfer simulation of reionization}
\label{section:2.3}

To post-process the simulation outputs with \texttt{CRASH}, we have gridded the gas density and temperature extracted from the snapshots onto $N_{\rm grid}^{3}$ cells.
More specifically, the RT is performed with $N_{\rm grid}=128$ (corresponding to a spatial resolution of 156 $h^{-1}\,\mathrm{ckpc}$) in the range $z=15-8$, while at $z<8$ the resolution is increased to $N_{\rm grid}=256$ (corresponding to 78 $h^{-1}\,\mathrm{ckpc}$) to ensure a better description of the Ly$\alpha$ forest related statistics. The gridded properties of the gas are obtained by assigning the particle data to a regular grid using the SPH kernel \citep{monaghan1992}, while the corresponding grid for the halo masses and positions is obtained by using the cloud-in-cell and friends-of-friends algorithms \citep{hockney1988,springel2001}. If more than one halo ends up in the same cell, their masses are combined. 
To reduce the large number of sources present in the simulation box and thus the computational time, we have adopted the source clustering technique developed by \citet{Marius2020} with a luminosity limit of $0.1$ and maximum radius of $\sqrt{2}\times 2.01$ (see the original paper for more details). Every ionizing source emits $10^{4}$ photon packets which are lost once they exit the simulation box, i.e. no periodic boundary condition has been used. 
\begin{figure} 
\centering
    \includegraphics[width=\columnwidth]{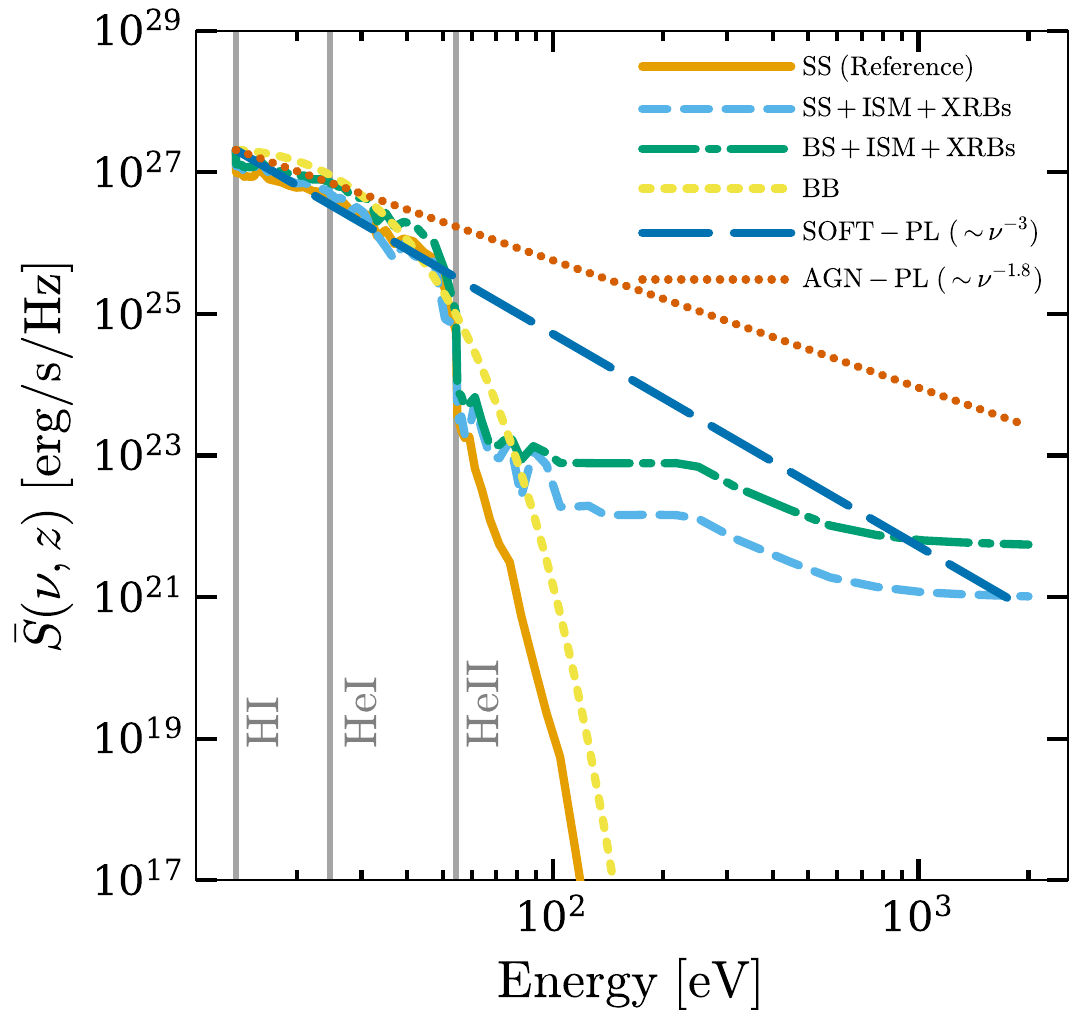} 
    \caption{Averaged spectral energy distribution $\bar{S}(\nu,z)$ for: single stars (\texttt{SS}, solid orange curve); a combination of single stars, ISM and XRBs (\texttt{SS+ISM+XRBs}, dashed sky blue); a combination of binary stars, ISM and XRBs (\texttt{BS+ISM+XRBs}, dash-dotted green); a black-body spectrum with 50,000 K (\texttt{BB}, short-dashed yellow); a power-law spectrum with a slope of -3 (\texttt{SOFT-PL}, long-dashed blue), and a power-law spectrum with a slope of -1.8 (\texttt{AGN-PL}, dotted vermilion). The vertical gray lines indicate the ionization thresholds for neutral hydrogen (13.6 eV), neutral helium (24.6 eV) and singly ionized helium (54.4 eV).
    }
    \label{fig:siminput_sed}
\end{figure}

For each output $i$ at $z_i$ of the hydrodynamical simulation, the RT is followed for a time $\rm{\mathit{t}_{H}(\mathit{z}_{i+1}) - \mathit{t}_{H}(\mathit{z}_{i})}$, where $\rm{\mathit{t}_{H}(\mathit{z}_{i})}$ is the Hubble time corresponding to $z_{i}$. 
In order to account for the expansion of the Universe between the $i$-th and $(i+1)$-th snapshots, the gas number density is evolved as $n(\mathbf{x},z) = n(\mathbf{x}, z_{i})(1 + z)^{3}/(1 + z_{i})^{3}$, where $\mathbf{x} \equiv (x_{c}, y_{c}, z_{c})$ are the coordinates of each cell $c$. The \texttt{Sherwood} simulations include a UV background (necessary to approximately capture the effect of radiation on gas dynamics as mentioned in Section \ref{hydrosim}) starting from $z=9$. Therefore, we need to remove its effects on the IGM temperature and ionization state, to avoid the double counting of ionizing photons when performing post-processing ionizing RT. Shocks and feedback processes can also heat up the gas, therefore uniformly resetting the gas temperature to some value would erase also such contribution. Since stellar radiation is not able to heat up the gas above $10^5$ K \citep{Draine1978,Barnes2014}, we assume that all gas particles hotter than that are affected primarily by shocks and feedback processes, and therefore are kept unchanged. For all other cells, we set their UVB-corrected temperature to be $\rm{\mathit{T}_{i+1}=\mathit{T}_{i}(1+\mathit{z}_{i+1}^{2})/(1+\mathit{z}_{i}^{2}}$). This amounts to assuming that they only undergo adiabatic cooling since the previous snapshot (since photo-ionization and photo-heating are included through the post-processing RT).
The ionization fractions are initialized to their expected residual values from the recombination epoch, i.e.  $x_{\rm HII}=10^{-4}$ and  $x_{\rm HeII}=x_{\rm HeIII}=0$.

\begin{figure*}
\centering
    \includegraphics[scale=0.385]{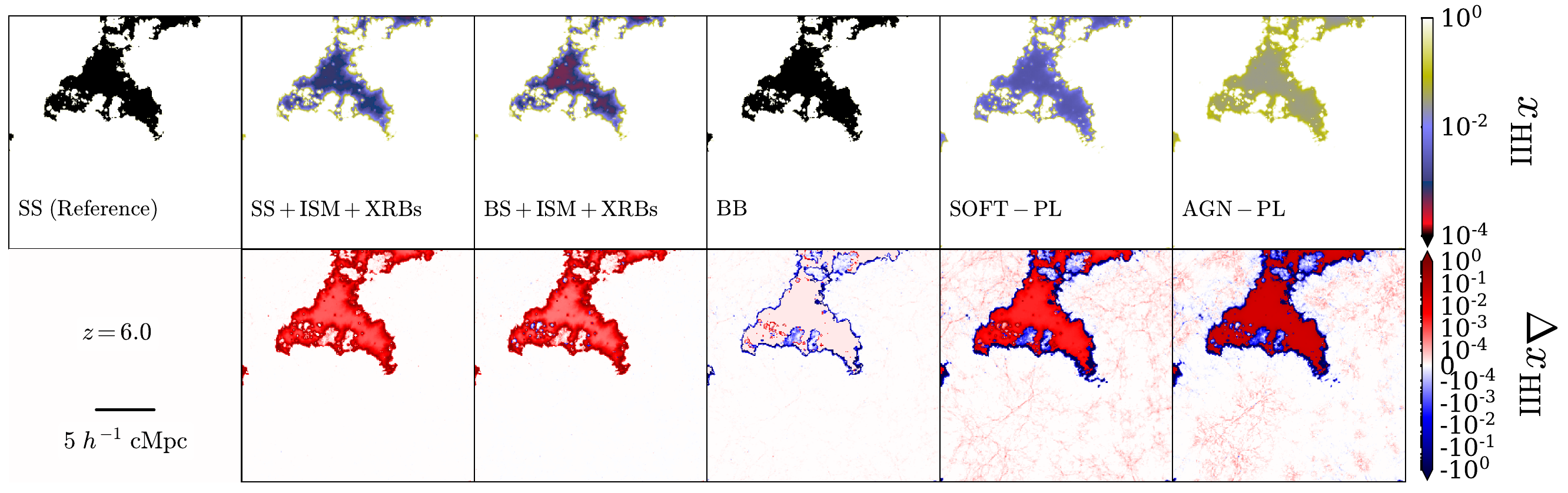}     
    \includegraphics[scale=0.385]{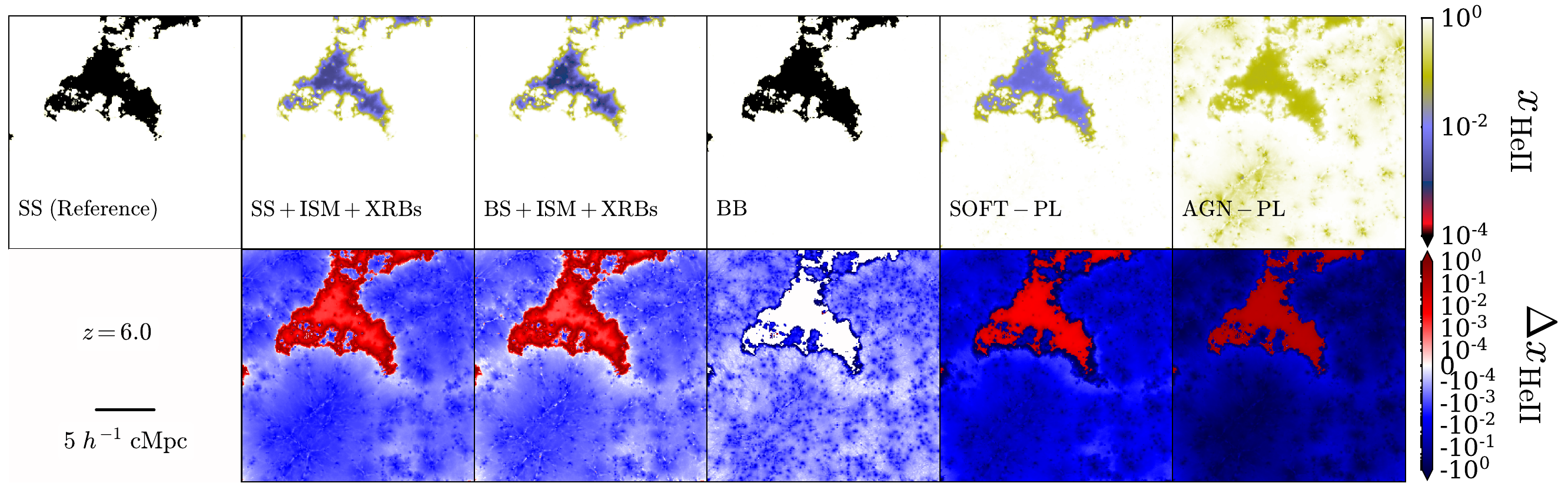}      
    \includegraphics[scale=0.385]{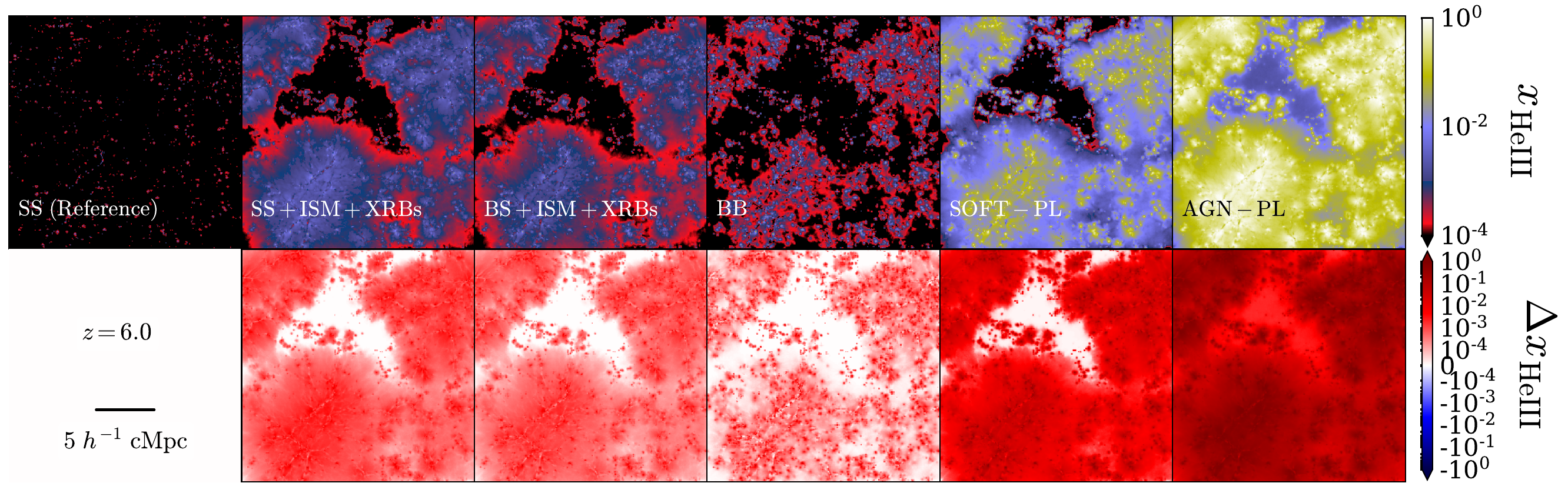}   
    \includegraphics[scale=0.385]{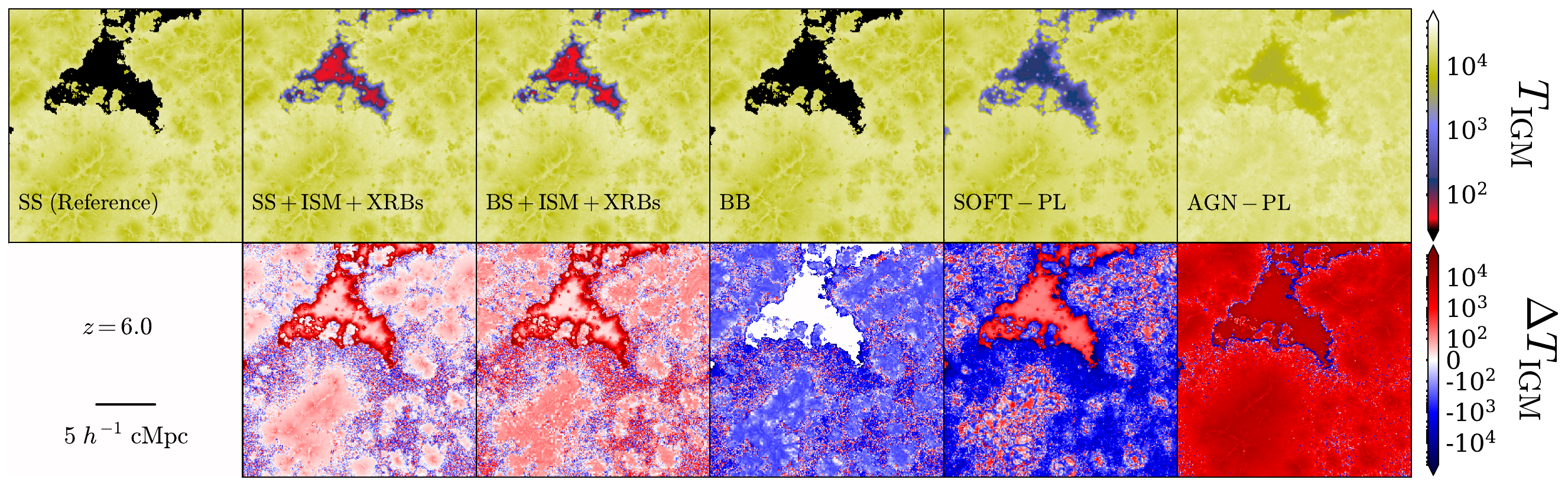}   
    \caption{From top to bottom, slice maps of the H~{\sc ii}, He~{\sc ii}, He~{\sc iii} fractions  and IGM temperature at $z = 6$ for different combinations of SED models: single stars  ({\tt SS}, first column);  single stars, ISM and XRBs ({\tt SS+ISM+XRBs}, second); binary stars, ISM and XRBs ({\tt BS+ISM+XRBs}, third); blackbody ({\tt BB}, fourth); galaxy power-law ({\tt SOFT-PL}, fifth) and AGN power-law ({\tt AGN-PL}, sixth). The lower sets of rows show the difference with respect to the simulations with single stars only. The maps are $20 h^{-1} \rm{cMpc}$ wide and $78 h^{-1} \rm{ckpc}$ thick.}
    \label{fig:slice_map}
\end{figure*}

Traditionally, models of ionizing photon production rate (or emissivity) in reionization studies focus on the cumulative contribution of ionizing photons from evolving stellar populations, often incorporating time-dependent source formation histories to estimate the global photon budget. However, in this study, we adopt a different approach. While we still utilize an ionizing photon production rate-based framework, our focus is not on the integrated luminosity evolution over time, but rather on how variations in the spectral energy distribution (SED) of the sources influence IGM and hence Ly$\alpha$ forest observables. By isolating the impact of SED shape, assuming a fixed ionizing photon production rate model, we aim to determine whether different assumptions about the source spectra can significantly alter interpretations of the Ly$\alpha$ forest. To this end, the source ionizing photon production rate is evaluated according to the prescription introduced by \citet{Ciardi2012}, namely, we have assumed a total comoving ionizing photon production rate at each redshift, $\epsilon\rm{_{tot}(\mathit{z}_i)}$, and distributed it in terms of rate of ionizing photons (${\dot{N}_{\text{ion}}}$) among all the halos  proportionally to their mass. We display our adopted redshift evolution of ${\dot{N}_{\text{ion}}}$ in Figure \ref{fig:siminput_emiss}. This approach assures by design that the  ionizing photon production rate is broadly consistent with $z= 5-6$ observations \citep{becker2021,gaikwad2023,davies2024}.
Additionally, this approach does not require any assumption on the escape fraction of ionizing photons, which is a very uncertain quantity (\citealt{vanzella2016,vanzella2018,matthee2018,naidu2018,steidel2018,fletcher2019}; for a review see \citealt{wood2000,Pratika2018}). Differently from \citet{Ciardi2012}, though, rather than having a purely linear relation between the ionizing ionizing photon production rate and the halo mass, we have incorporated Gaussian noise scatter with mean $\mu=0$ and standard deviation $\sigma=0.01$ in the log-log scale, conserving the total ionizing photon production rate (as done in \citealt{enrico2019,Basu2024}). The adopted scatter is intended to represent modest halo-to-halo variability in ionizing output, and we have tested that it has a negligible impact on the large-scale Ly$\alpha$ forest statistics analyzed here, which are primarily sensitive to the ensemble-averaged emissivity.

The SED of the sources has been derived following \citet{Marius2018}, \citet{Marius2020} and \citet{qingbo2022}, which include the contribution from single (SS) and binary (BS) stars, XRBs and thermal Bremsstrahlung from supernova-heated ISM. While we refer the reader to the original papers for more details, here we note that stars dominate the total SED budget at energy $\rm{\mathit{h}_p \nu \leq 60}$ eV, while the ISM contribution is dominant above that into hard UV and the soft X-rays (i.e. in the range $\rm{\mathit{h}_p \nu \in [60,500]}$ eV),  and the XRBs start to contribute  at $\rm{\mathit{h}_p \nu > 500}$ eV (see Figure 2 in \citet{Marius2018} and Figure 1 in \citet{qingbo2022} for a visual inspection of individual SEDs of different types of ionizing sources). 
For all RT simulations, the spectra of ionizing sources extend to a maximum frequency of $\rm{\sim 2 \ keV}$ and are discretized into 82 frequency bins, with  bins more densely spaced around the ionization thresholds of hydrogen (13.6 eV) and helium (24.6 eV and 54.4 eV). We run three  simulations: one with ionizing sources consisting solely of single stars (\texttt{``SS"}), another adding also the contribution from the ISM and XRBs (\texttt{``SS+ISM+XRBs"}), and a third that includes binary stars along with the ISM and XRBs (\texttt{``BS+ISM+XRBs"}). For comparison, we also perform three additional simulations using simplified SED prescriptions commonly adopted in the literature. These include a blackbody spectrum with $T$ = 50,000 K (\texttt{``BB"}), a galaxy-like power-law SED with a slope of -3 (\texttt{``SOFT-PL"}), and an AGN-type power-law SED with a slope of -1.8 (\texttt{``AGN-PL"}). For simplicity, following \citet{Marius2018} and \citet{Marius2020}, we adopt the globally averaged SEDs (averaging over the spectral shapes for all sources) for the first three simulations. For the last three simulations, we assume a constant model for all sources. The resulting SEDs for all six simulations are shown in Figure \ref{fig:siminput_sed}.

\section{Results}
\label{results}

The central question we seek to explore is: \textit{what is the impact of SED models on the properties of the IGM as revealed by the Ly$\alpha$ forest?} To address this, we will first examine the qualitative differences introduced by different SEDs of sources and the resulting photoionization rate, then delve into the quantitative details of reionization and thermal histories. Finally, we will analyze their influence on the observable features associated with the Ly$\alpha$ forest.

\subsection{Qualitative overview of IGM properties}
\label{slice_compare}

For a qualitative discussion of the results, we present the maps of the ionization and thermal state of the IGM at $z = 6$ for all simulations in Figure \ref{fig:slice_map}, together with maps showing the difference relative to \texttt{SS}.

In the \texttt{SS} model, the gas is mostly either fully ionized and hot ($x_{\rm HII} \approx x_{\rm HeII} \approx 1$, $T_{\rm IGM} \gtrsim 10^4$ K) or nearly neutral and cold ($x_{\rm HII} \approx x_{\rm HeII} \lesssim 10^{-4}$, $T_{\rm IGM} \sim 10$ K). 
In contrast, the higher-energy photons emitted by the ISM and XRBs in the \texttt{SS+ISM+XRBs} model introduce extra heating and partial ionization outside of the fully ionized regions, raising the gas temperature by few hundred degrees. 
This is evident in the difference maps of H~{\sc ii}, He~{\sc ii}, and $\rm{\mathit{T}_{IGM}}$, where red pixels highlight regions with increased ionization and heating. He~{\sc ii} behaves similarly to H~{\sc ii}, with the exception that within fully ionized regions, He~{\sc ii} is partially converted into He~{\sc iii}, as it is clearly visible from both maps and difference maps. The impact of He~{\sc iii} is also seen as an increment in the temperature of the ionized regions. We also note that the noise-like pattern in the $\rm{\mathit{T}_{IGM}}$ difference maps within fully ionized regions results from Monte Carlo noise, as obtaining exactly the same temperature between simulations with different ionizing sources is statistically unlikely \footnote{We note that the difference is well within $\sim 1-2\%$.}. However, this is not the case for the ionization fraction once a cell is fully ionized. 
The inclusion of binary stars in the \texttt{BS+ISM+XRBs} simulation displays a very similar ionization and thermal structure of the IGM as \texttt{SS+ISM+XRBs}. However many of the higher energy photons are used in lifting the level of partial ionization surrounding the fully ionized regions, resulting in fewer photons available to penetrate further into the neutral IGM, and leading to reduced partial ionization in more distant regions. The \texttt{BS+ISM+XRBs} model results in a slightly lower He~{\sc iii} fraction than \texttt{SS+ISM+XRBs} because a significant fraction
of ionizing photons are consumed in expanding the ionized bubbles, while the total ionizing photon budget per time step remains fixed.

While comparing with the simplified SED prescriptions commonly adopted in literature for reionization studies, we notice certain differences with respect to our reference \texttt{SS} model. The extent of the fully ionized regions (i.e. $x_{\rm HII} \gtrsim 0.99$) in the \texttt{BB} model is slightly smaller than in the reference \texttt{SS} one, as seen from the dark blue pixels surrounding the ionized gas in the difference maps of $x_{\rm HII}$ and $x_{\rm HeII}$. This is primarily due to a slight difference in the numbers of photons around the H~{\sc i} ionization threshold; otherwise, the maps in the \texttt{BB} model are qualitatively the same as in the \texttt{SS} model. However, the amount of doubly ionized helium is higher in \texttt{BB}, since this model produces more photons at and slightly beyond the He~{\sc ii} ionization energy compared to the \texttt{SS} model. Since the total ionizing photon production rate is fixed across all models, a higher abundance of photons in the high-frequency regime necessarily reduces the relative number of photons near the H~{\sc i} ionization threshold. Thus, the \texttt{BB} model, which produces more photons around and beyond the He~{\sc ii} ionization threshold, contains comparatively fewer H~{\sc i} ionizing photons. 
As a result, the reduced hydrogen photoheating leads to a slightly lower IGM temperature than in the \texttt{SS} model, despite the additional (but subdominant) heating contributed by the ionization of helium.
For similar reasons, in the \texttt{SOFT-PL} model the extent of the fully ionized regions is even smaller.
The \texttt{SOFT-PL} model also has a higher fraction of gas partially ionized by the higher energy tail of the SED, with many cells reaching $x_{\rm HII} \sim 10^{-2}$. The temperature maps further reflect this, showing increased heating in the partially ionized regions and inside ionized bubbles, a similar but more pronounced feature as in the \texttt{BB} model. However, in the dense gas, additional helium double ionization heating becomes the dominant factor in raising the temperature. The difference maps in $x_{\rm HII}$ reveal filamentary structures with positive $\Delta x_{\rm HII}$, representing regions which have lower recombination rates due to a higher temperature. 
In the \texttt{AGN-PL} model, all the effects discussed above are even more pronounced. The higher number of high-energy photons leads to stronger partial ionization of H~{\sc i} and He~{\sc i}, with many cells reaching $x_{\rm HII}$ and $x_{\rm HeII} \sim 10^{-1}$. These energetic photons also drive more helium double ionization, creating large regions of fully ionized helium. As a consequence, the overall temperature is significantly higher, and the temperature difference maps appear mostly red, reflecting widespread heating across the slice. 

While these differences highlight the qualitative impact of various SED model assumptions, a more insightful approach is to perform a quantitative analysis. In the following sections, we will focus on this aspect.

\subsection{H~{\sc i} photoionization rate}
\label{photoionizationrate}
\begin{figure}   
\centering
\includegraphics[width=\columnwidth]{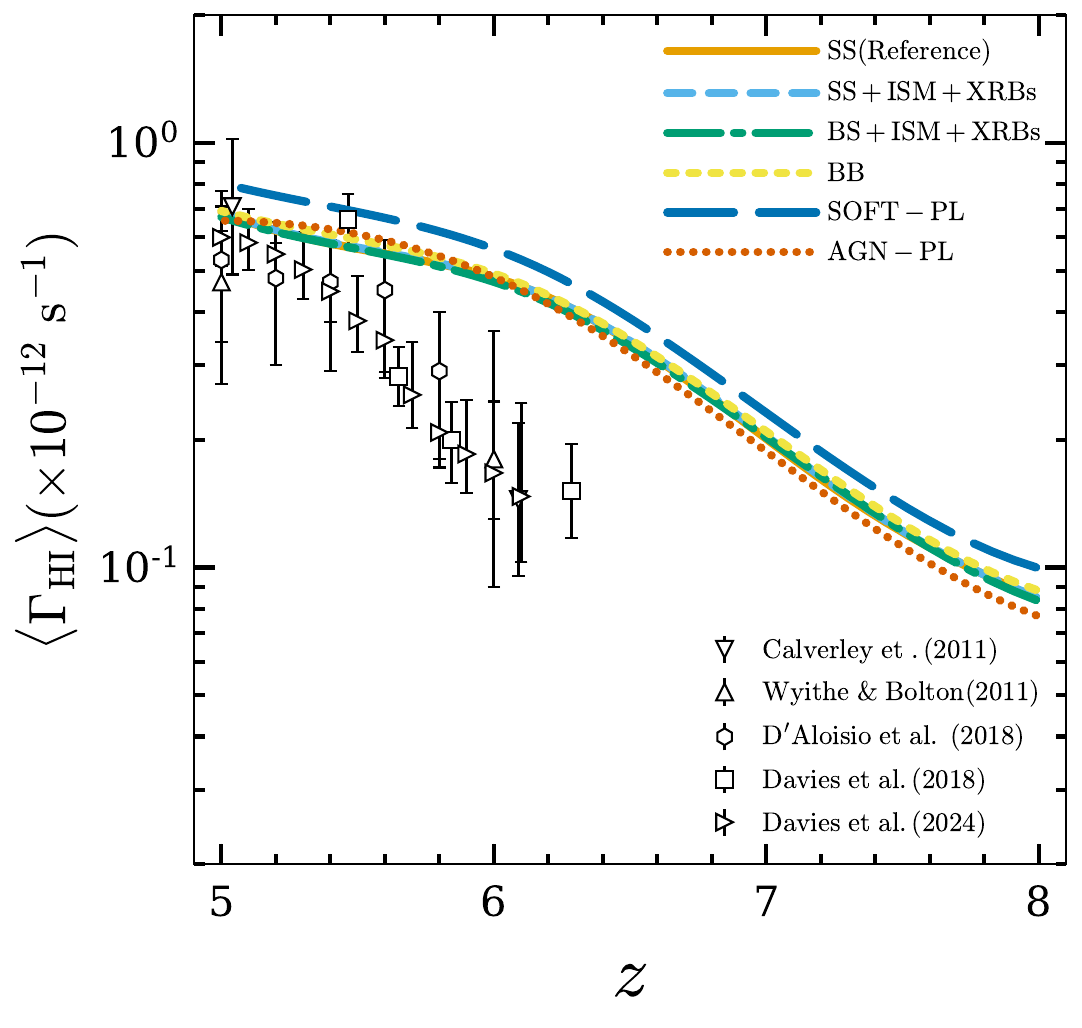}   
    \caption{Volume averaged H~{\sc i} photoionization rate for model \texttt{SS} (solid orange curve), \texttt{SS+ISM+XRBs} (dashed sky blue), \texttt{BS+ISM+XRBs} (dash-dotted green), \texttt{BB} (short-dashed yellow), \texttt{SOFT-PL} (long-dashed blue) and \texttt{AGN-PL} (dotted vermilion). The symbols denote a compilation of observational constraints from the literature \citep{Calverley2011,wyithe2011,daloisio2018,Davies2018, davies2024}.}
    \label{fig:gammahi}
\end{figure}
In Figure \ref{fig:gammahi}, we present the volume averaged H~{\sc i} photoionization rate $\rm{\langle \mathit{\Gamma}_{HI} \rangle}$ extracted from our simulations at $z \lesssim 8$, where we achieve a higher spatial resolution for the RT calculations (as discussed in Section \ref{section:2.3}). 
Since the photoionization rate is not a direct output of our simulations, we estimate it indirectly by assuming ionization equilibrium in each simulation cell (following the approach of \citealt{Ciardi2012}). Under this assumption, the rate can be expressed as:
\begin{equation}
\Gamma_{\mathrm{HI}} = \alpha_{\mathrm{HII}}(T) \frac{n_e n_{\mathrm{HII}}}{n_{\mathrm{HI}}} - \gamma_{e\mathrm{HI}}(T)n_e,
\end{equation}
where $\alpha_{\mathrm{HII}}(T)$ and $\gamma_{e\mathrm{HI}}(T)$ are the temperature-dependent hydrogen recombination and collisional ionization rates (in units of cm$^3$\,s$^{-1}$), respectively. All other quantities have their standard physical meanings. This equilibrium-based estimate is generally valid for most of the cells in our simulation volume after reionization. 
The results are shown for all simulations, alongside observational constraints from \citealt{Calverley2011}, \citealt{wyithe2011}, \citealt{daloisio2018}, \citealt{Davies2018} and \citealt{davies2024}. Note that we have calculated  $\rm{\langle \mathit{\Gamma}_{HI} \rangle}$ taking into account only the ionized cells (with $x_{\rm HII} \gtrsim 0.99$), although we notice that computing $\rm{\langle \mathit{\Gamma}_{HI} \rangle}$ from all cells in the simulation volume shows marginal differences (i.e. deviation within 5 $\%$ at $z\lesssim7$). Nevertheless, all simulations follow a similar overall trend, with $\rm{\langle \mathit{\Gamma}_{HI} \rangle}$ increasing by approximately an order of magnitude from $z = 8$ to $z = 5$. At $z \sim 6$, the evolution of $\rm{\langle \mathit{\Gamma}_{HI} \rangle}$ slows down and eventually flattens, marking the phase when the majority of the simulation volume becomes ionized. 

Although most simulations yield photoionization rates that are systematically higher than the observed ones, there are distinct, albeit minor, differences between the models. When comparing the different models to the \texttt{SS} reference simulation, we observe that despite the general trend being similar, differences emerge. Notably, the \texttt{SOFT-PL} model overall yields the largest $\rm{\langle \mathit{\Gamma}_{HI} \rangle}$ value, higher by a factor of $\sim 1.15$ compared to \texttt{SS}. This is primarily due to the lower IGM temperature, which results in a higher recombination rate, and hence H~{\sc i} photoionization rate. Meanwhile, models with other SEDs, i.e. \texttt{BB}, \texttt{SS+ISM+XRBs}, and \texttt{BS+ISM+XRBs} exhibit similar trends, with \texttt{BB} producing the highest $\rm{\langle \mathit{\Gamma}_{HI} \rangle}$ among them for the same reason.
The behavior of the \texttt{AGN-PL} model is particularly noteworthy. At the highest redshifts, it shows the lowest $\rm{\langle \mathit{\Gamma}_{HI} \rangle}$, largely because the higher IGM temperatures (as discussed in Fig.~\ref{fig:slice_map}) suppress the recombination rate. However, for $z < 6$ the trend reverses: $\rm{\langle \mathit{\Gamma}_{HI} \rangle}$ rises sharply and overtakes all other models (except \texttt{SOFT-PL}), mostly because of the dominant effect of the enhanced supply of free electrons from the pronounced double ionization of helium.
In the following sections we conduct a detailed analysis of IGM ionization and heating, comparing various simulations, and then explore the impact on the characteristics of the Ly$\alpha$ forest. 

\begin{figure}    \includegraphics[width=\columnwidth]{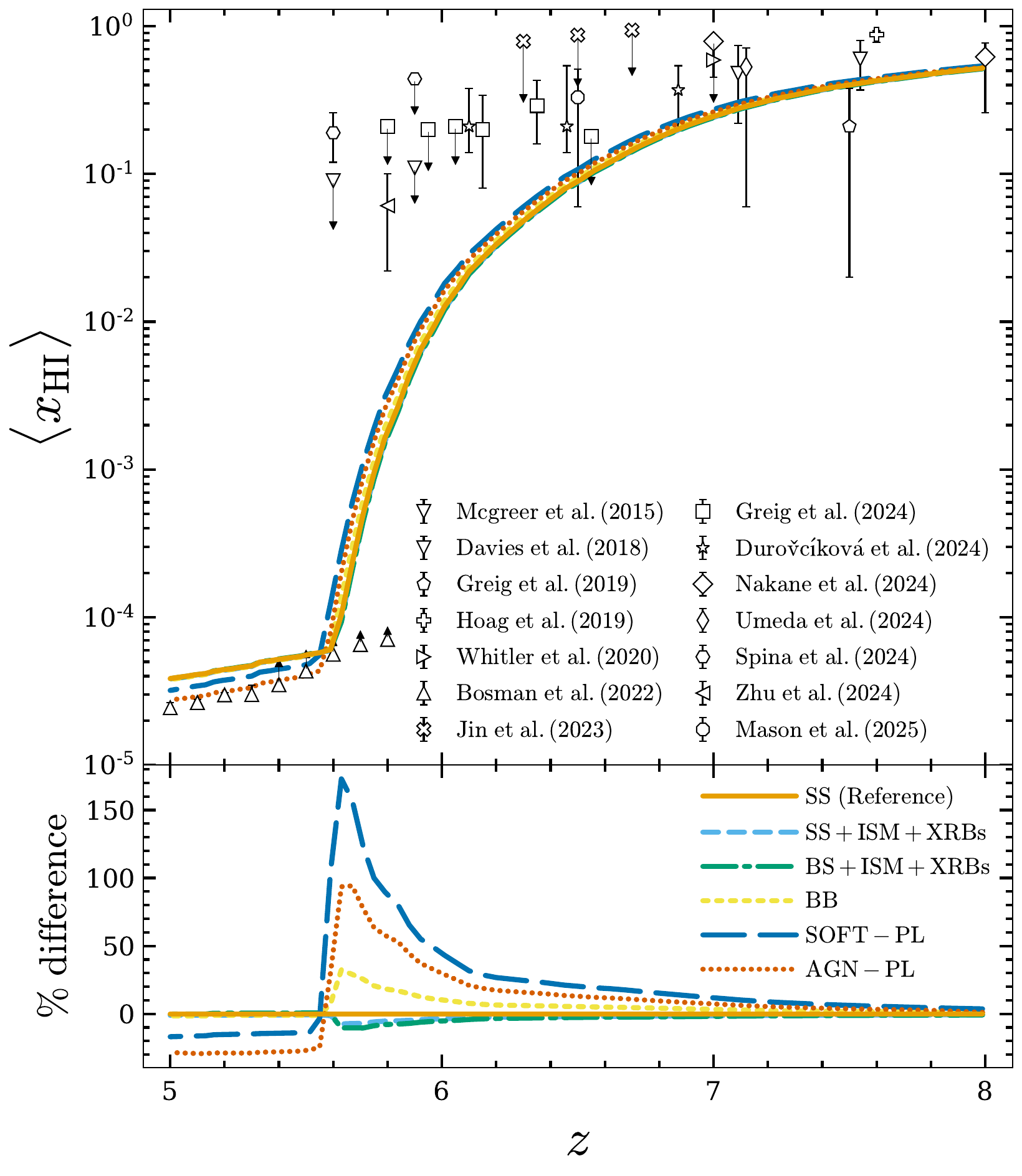}   
    \caption{{\it Top panel:} Redshift evolution of the volume averaged $\mathit{x}\rm{_{HI}}$ for all SED models: \texttt{SS} (solid orange curve), \texttt{SS+ISM+XRBs} (dashed sky blue), \texttt{BS+ISM+XRBs} (dash-dotted green), \texttt{BB} (short-dashed yellow), \texttt{SOFT-PL} (long-dashed blue) and \texttt{AGN-PL} (dotted vermilion). The symbols denote a compilation of observational constraints from the literature \citep{McGreer2015,Davies2018,greig2019,hoag2019,Whitler2020,Bosman2022,Jin2023,Greig2024,durovcicova2024,nakane2024,umeda2024,spina2024,zhu2024,Mason2025}. {\it Bottom panel:} Fractional difference with respect to the \texttt{SS} simulation.}
    \label{fig:XHIevol}
\end{figure}

\subsection{Reionization history}
\label{sec:reion}
To discuss the overall effect induced by different SED models, in Figure \ref{fig:XHIevol} we display the redshift evolution of the volume averaged H~{\sc i} fraction for all simulations, together with the fractional difference with respect to \texttt{SS}, and a compilation of observational results \citep{McGreer2015,Davies2018,greig2019,hoag2019,Whitler2020,Bosman2022,Jin2023,Greig2024,durovcicova2024,nakane2024,umeda2024,spina2024,zhu2024,Mason2025}. With the exception of \texttt{AGN-PL}, all simulations produce a neutral hydrogen fraction which is slightly higher than observational constraints at $z \lesssim 5.5$, where our primary focus lies. The curves are able to reproduce most data points at higher redshift, where though there is less consistency among different observations.  

As discussed earlier, the inclusion of binary stars or more energetic sources leads to increased partial ionization and heating, resulting in a lower H~{\sc i} fraction. The most significant differences appear at $z = 5.6$, where the \texttt{SS+ISM+XRBs} and \texttt{BS+ISM+XRBs} models show deviations of approximately 7.5$\%$ and 10$\%$, respectively. In contrast, the \texttt{BB} model maintains a higher H~{\sc i} fraction up to $z = 5.6$, indicating a slightly delayed reionization (by about $\Delta z \sim 0.05$) compared to the \texttt{SS} model. This delay is even more pronounced in the \texttt{SOFT-PL} and \texttt{AGN-PL} models, which sustain higher H~{\sc i} fractions over a similar redshift range. This occurs because, as the total number of ionizing photons is kept constant and these models contain more high-energy photons, they have fewer in the vicinity of the H~{\sc i} ionization potential, which are more efficient at ionizing H~{\sc i} (as discussed in Section \ref{slice_compare}). The \texttt{SOFT-PL} model shows the largest deviation, with a fractional difference exceeding 170$\%$ at $z = 5.6$, while the deviation for the \texttt{AGN-PL} model reaches about 100$\%$.
As reionization progresses and $\langle x_{\rm HI} \rangle$ falls below $10^{-4}$, the differences between the models diminish and eventually vanish. However, during the final stages of reionization, the \texttt{SOFT-PL} and \texttt{AGN-PL} models predict a Universe that is more ionized than in the \texttt{SS} model, with deviations of approximately 15$\%$ and 30$\%$, respectively—primarily due to additional heating and its impact on the recombination rate, as discussed in Section \ref{photoionizationrate}.

\begin{figure}    \includegraphics[width=\columnwidth]{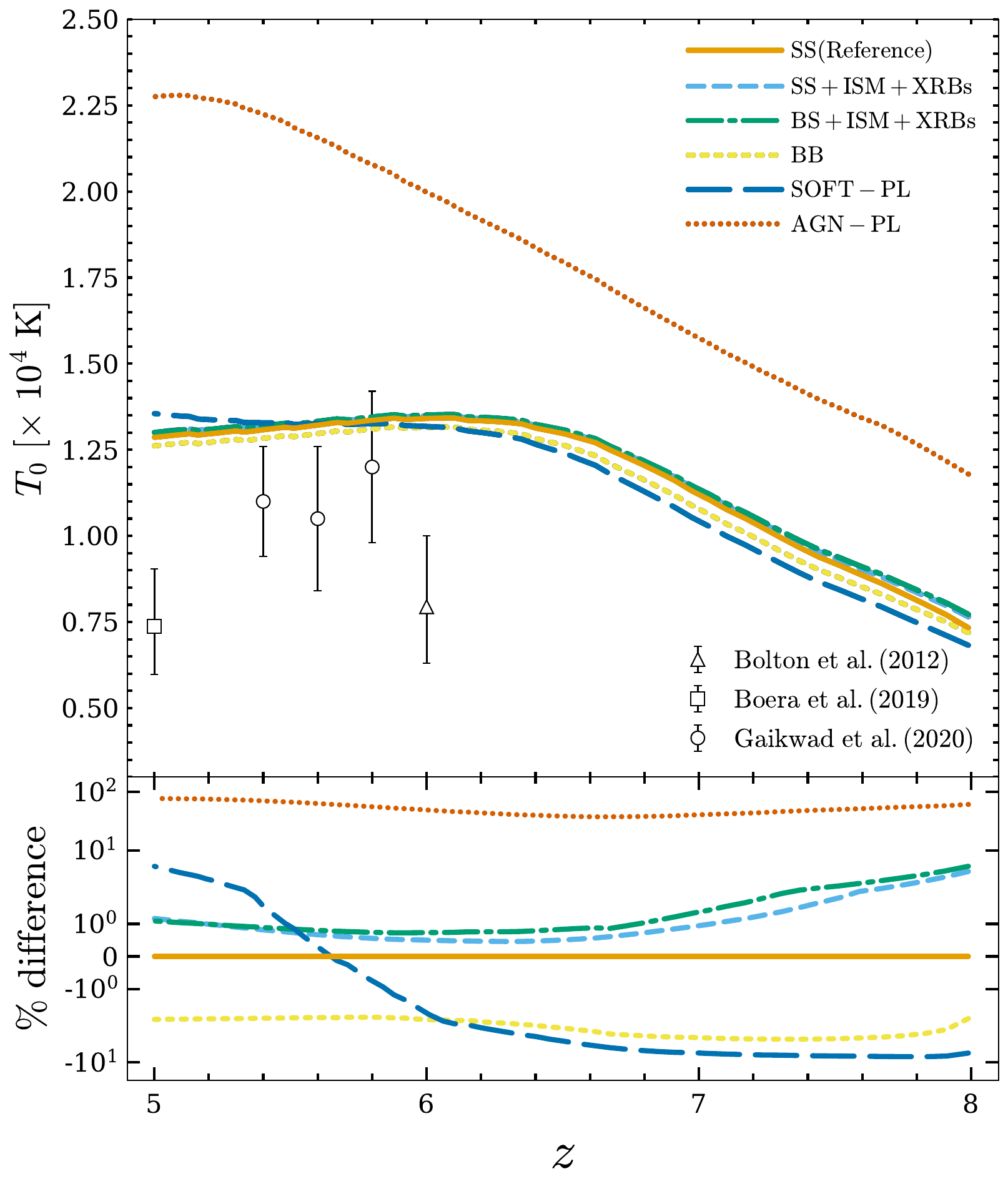}   
    \caption{{\it Top panel:} Redshift evolution of the IGM temperature at mean density ($\mathit{T}\rm{_{0}}$) for \texttt{SS} (solid orange curve), \texttt{SS+ISM+XRBs} (dashed sky blue), \texttt{BS+ISM+XRBs} (dash-dotted green), \texttt{BB} (short-dashed yellow), \texttt{SOFT-PL} (long-dashed blue) and \texttt{AGN-PL} (dotted vermilion). The symbols denote a compilation of observational constraints from the literature \citep{Bolton2012,boera2019,Gaikwad2020}. {\it Bottom panel:} Fractional difference with respect to the \texttt{SS} simulation.}
    \label{fig:tempevol}
\end{figure}

\subsection{Thermal history}
\label{thermal_history}
The ionization of hydrogen and helium injects energy into the IGM through photoheating, raising its temperature, and consequently affecting the dynamics of gas, its cooling rate, and its accretion onto galaxies. In Figure \ref{fig:tempevol}, we show the redshift evolution of the IGM temperature at mean density, calculated by selecting all the cells in the simulation volume with gas density within $\pm10\%$ of the mean density.
As expected, with the exception of \texttt{AGN-PL}, $ T_{0} $ rises with decreasing redshift, reaching a peak at $ z= 6 $, when about 99\% of the simulation volume is ionized. Subsequently, the rate of heat injection from photoheating diminishes, leading to a plateau followed by a gradual decline, which is much shallower, though, than the one  suggested by a compilation of observational data \citep{Bolton2012,boera2019,Gaikwad2020}. A similar behavior has been observed in the context of helium reionization by \citet{Basu2024}. 
The apparent discrepancies at $z>5$ should be interpreted with caution, as the available observational constraints may be affected by systematic uncertainties that introduce scatter. In addition, because these constraints are inferred from the Ly$\alpha$ forest, they exclude contributions from very hot, shock-heated regions of the IGM. However, we have verified that the shallower evolution seen in our simulations is not driven by additional heating from the increasing number of cells affected by hydrodynamical feedback. We have explored this point quantitatively by computing the thermal history taking into account only the cells below a temperature threshold of $2 \times 10^{4}$ K, $5 \times 10^{4}$ K and $10^{5}$ K, finding maximum differences of $3\%$  with respect to the case when all cells are considered.
While the redshift evolution of the \texttt{SOFT-PL} model slows at $z=6$, the temperature continues to rise at lower redshifts, primarily due to the effects of helium double ionization. A similar trend is evident in the \texttt{AGN-PL} model, though with a significantly higher amplitude, consistently exceeding the temperatures of other models by more than 7500 K due to the abundant energetic photons populating this SED. Interestingly, though, while the evolution of the  H~{\sc i} fraction is mostly consistent with observations, in particular at $z<5.5$, the IGM temperature is always substantially higher than the data points.  

Both \texttt{SS+ISM+XRBs} and \texttt{BS+ISM+XRBs} exhibit temperatures higher than \texttt{SS} due to additional heating from more energetic photons, with binary stars contributing even more significantly. The maximum temperature difference reaches approximately $5\%$ at $z = 8$, but gradually decreases to about $1\%$ at lower redshifts, as reionization progresses.  
The \texttt{BB} model consistently shows lower temperatures, with a nearly constant deviation of $\sim 4\%$ for $z \lesssim 6$ as discussed in Section \ref{slice_compare}. Differently, \texttt{SOFT-PL}  initially produces temperatures lower than the \texttt{SS} model, reaching a maximum negative deviation of $\sim 10\%$ at $z = 8$. 
However, below $z = 6$, with most of the H~{\sc i} and He~{\sc i} in the simulation volume getting ionized, this deviation gradually decreases and eventually turns positive, reaching approximately $8\%$ at $z = 5$ due to a more pronounced effect from double helium reionization. 
As already mentioned, the \texttt{AGN-PL} model is characterized by larger values of $T_0$ at all redshifts, with deviations of $\sim 100\%$.

In the following section, we will investigate if the differences noticed in IGM temperature and ionization state affect the Ly$\alpha$ forest properties.

\subsection{Ly$\alpha$ forest statistics}
\label{lyalpha_forest}
To investigate Ly$\alpha$ forest properties, we have generated synthetic spectra by extracting 16384 sightlines at each RT snapshot, each spanning the full box length parallel to the $z$-direction. For each pixel $i$ in a sightline we evaluate the normalized transmitted flux $F(i) = \exp[-\tau(i)]$, where
\begin{equation}
    \tau(i) = \frac{c \sigma_{\alpha} \delta R}{\sqrt{\pi}} \sum_{j=1}^{N} \frac{n\mathrm{_{HI}}(j)}{b\mathrm{_{HI}}(j)} \ \tilde{V}(i,j)
\end{equation}
is the Ly$\alpha$ optical depth. Here $N$ is the number of pixels in a sightline, $\sigma_{\alpha}$ is the Ly$\alpha$ scattering cross-section, $c$ is the speed of light, $\delta R$ is the size of a pixel in proper distance units, $n_\mathrm{{HI}}(j)$ is the H~{\sc i} number density at the position of pixel $j$, $\mathrm{\mathit{b}_{HI}(\mathit{j}) = (2 \mathit{k}_{B} \mathit{T}_{IGM} / \mathit{m}_{H})^{1/2}}$, $\mathrm{\mathit{k}_{B}}$ is the Boltzmann constant, $\mathrm{\mathit{m}_{H}}$ is the hydrogen mass, and $\tilde{V}$ is the Voigt profile approximation provided by \citet{tepper2006}. The latter depends on the IGM temperature and the peculiar velocity. Throughout this paper, we ignore the contamination from other lines into the wavelength window of the Ly$\alpha$ forest as this is expected to have a negligible
impact on our results \citep{hellsten1998,yang2022}. 

\begin{figure}
\centering
    \includegraphics[width=\columnwidth]{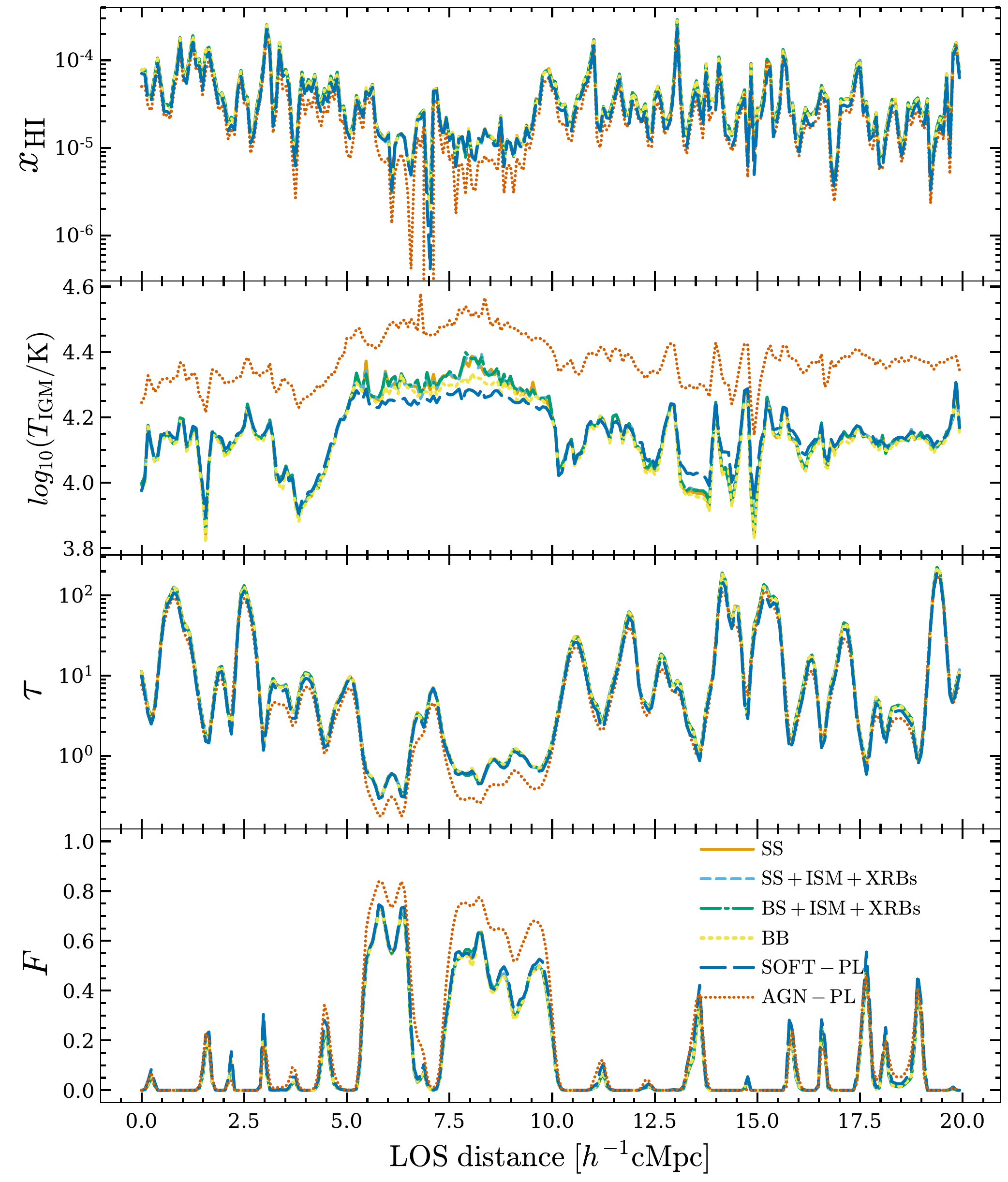}
    \caption{IGM properties along a reference line of sight at \textit{z} = 5 for \texttt{SS} (solid orange curve), \texttt{SS+ISM+XRBs} (dashed sky blue), \texttt{BS+ISM+XRBs} (dash-dotted green), \texttt{BB} (short-dashed yellow), \texttt{SOFT-PL} (long-dashed blue) and \texttt{AGN-PL} (dotted vermilion). From top to bottom the panels refer to the H~{\sc i} fraction, IGM temperature, Ly$\alpha$ optical depth, and normalized transmitted flux. }
    \label{fig:los}
\end{figure}

\subsubsection{Individual lines of sight}
\label{los}

As a reference, in Figure \ref{fig:los} we display the neutral hydrogen fraction ($\rm{\mathit{x}_{HI}}$), the IGM temperature ($\rm{\mathit{T}_{IGM}}$), the Ly$\alpha$ optical depth ($\rm{\tau}$), and the Ly$\alpha$ transmitted flux ($F$) along a single line of sight (LOS) at $z=5$. For clarity, we omit the pixel index when referring to the values of these physical quantities.

The abundance of H~{\sc i} remains remarkably consistent across all simulations, with only slight variations. Notably, a marginally lower value is observed in the {\tt AGN-PL} model, as illustrated also in Figure~\ref{fig:XHIevol}. 
The ionized regions are characterized by an IGM temperature $T_{\rm IGM} \sim 10^4$~K, with  {\tt AGN-PL} exhibiting the highest value of $\sim 10^{4.4}$~K, as discussed in previous sections. In all models the highest temperatures are reached in regions which have been recently ionized. 
The behaviour observed in the H~{\sc i} fraction is directly reflected in the corresponding Ly$\alpha$ optical depth, which has typical values below 5 in regions of higher ionization. It is in these areas that we also observe a larger transmission of Ly$\alpha$ photons. When comparing different SED models, the {\tt AGN-PL} consistently exhibits the highest transmission, suggesting that its additional heating affects the width and shape of transmission features as a result of the lower recombination rate due to the higher temperature.
Instead, the other models do not show substantial variations in the transmitted flux, as the differences in optical depth introduce only subtle fluctuations primarily driven by the additional heating. This slightly enhances the transmission by reducing small-scale variations in optical depth. 

For a better comparison, we have computed the mean transmission flux ($\rm{\bar{\mathit{F}}}$) for the entire simulation volume at $z=5$. Most models, i.e. {\tt SS}, {\tt SS+ISM+XRBs}, {\tt BS+ISM+XRBs} and {\tt BB}, have $\rm{\bar{\mathit{F}}} \sim 0.14$. The slightly higher IGM temperature in the {\tt SOFT-PL} model results in $\rm{\bar{\mathit{F}}} \sim 0.16$. This difference becomes even more pronounced in the {\tt AGN-PL}, which has a $\sim 30 \%$ increase in the mean transmission flux in comparison to the {\tt SS} model, and $\rm{\bar{\mathit{F}}}=0.18$. 

While the characteristics discussed above are common to the majority of lines of sight, to provide a more quantitative comparison, in the next sections we calculate quantities which more accurately capture the statistical behaviour of the Ly$\alpha$ forest.

\begin{figure}   
\includegraphics[width=\columnwidth]{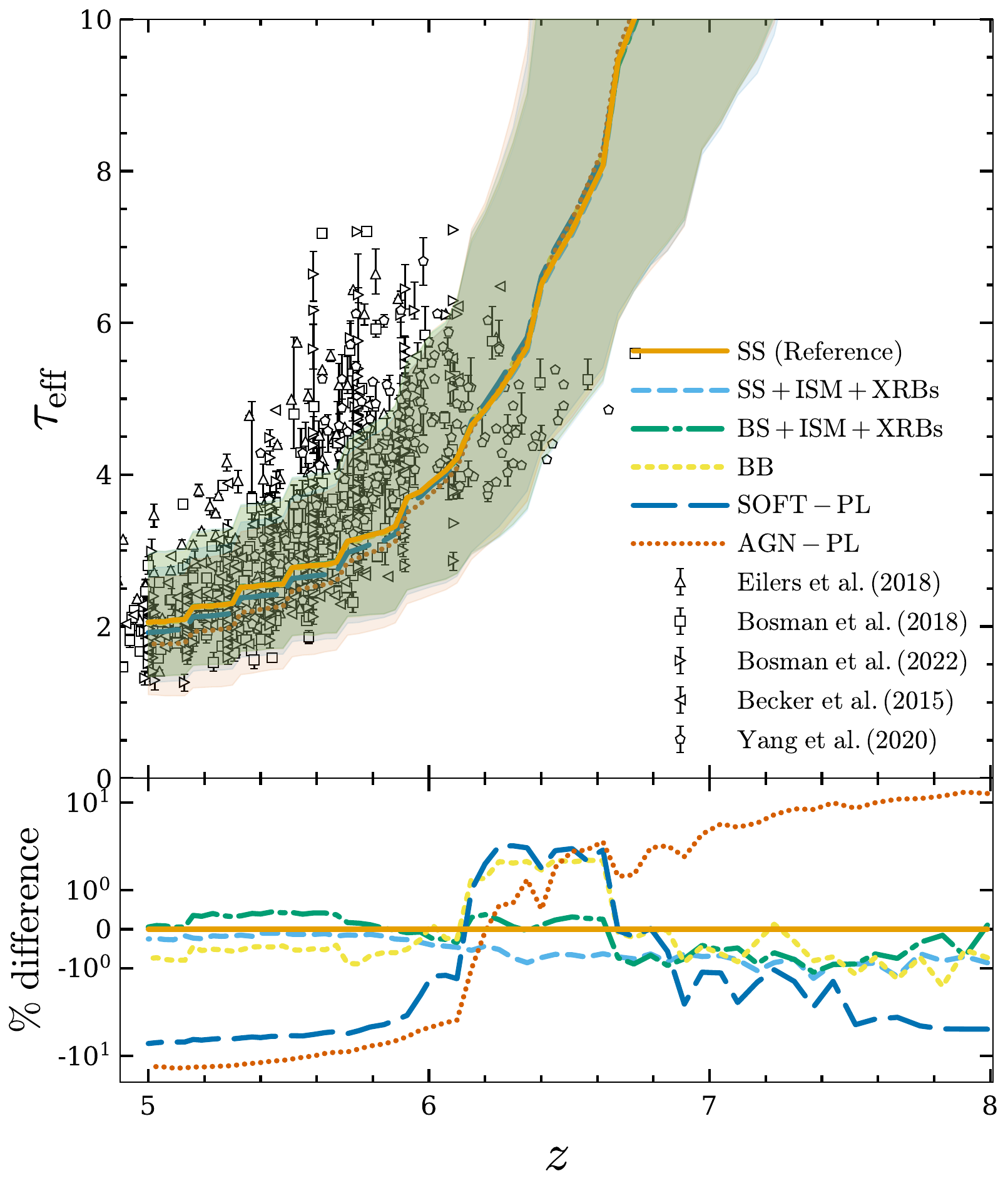}
    \caption{{\it Top panel:} Redshift evolution of the median Ly$\alpha$ effective optical depth for \texttt{SS} (solid orange curve), \texttt{SS+ISM+XRBs} (dashed sky blue), \texttt{BS+ISM+XRBs} (dash-dotted green), \texttt{BB} (short-dashed yellow), \texttt{SOFT-PL} (long-dashed blue) and \texttt{AGN-PL} (dotted vermilion). The shaded regions indicate 95$\%$ confidence intervals. The symbols denote a compilation of observational constraints from the literature \citep{Becker2015,eilers2018,yang2020,bosman2018,Bosman2022}. {\it Bottom panel:} Fractional difference with respect to the \texttt{SS} simulation. }
    \label{fig:taueff_evol}
\end{figure}

\begin{figure}   
\centering
\includegraphics[width=1.05\columnwidth]{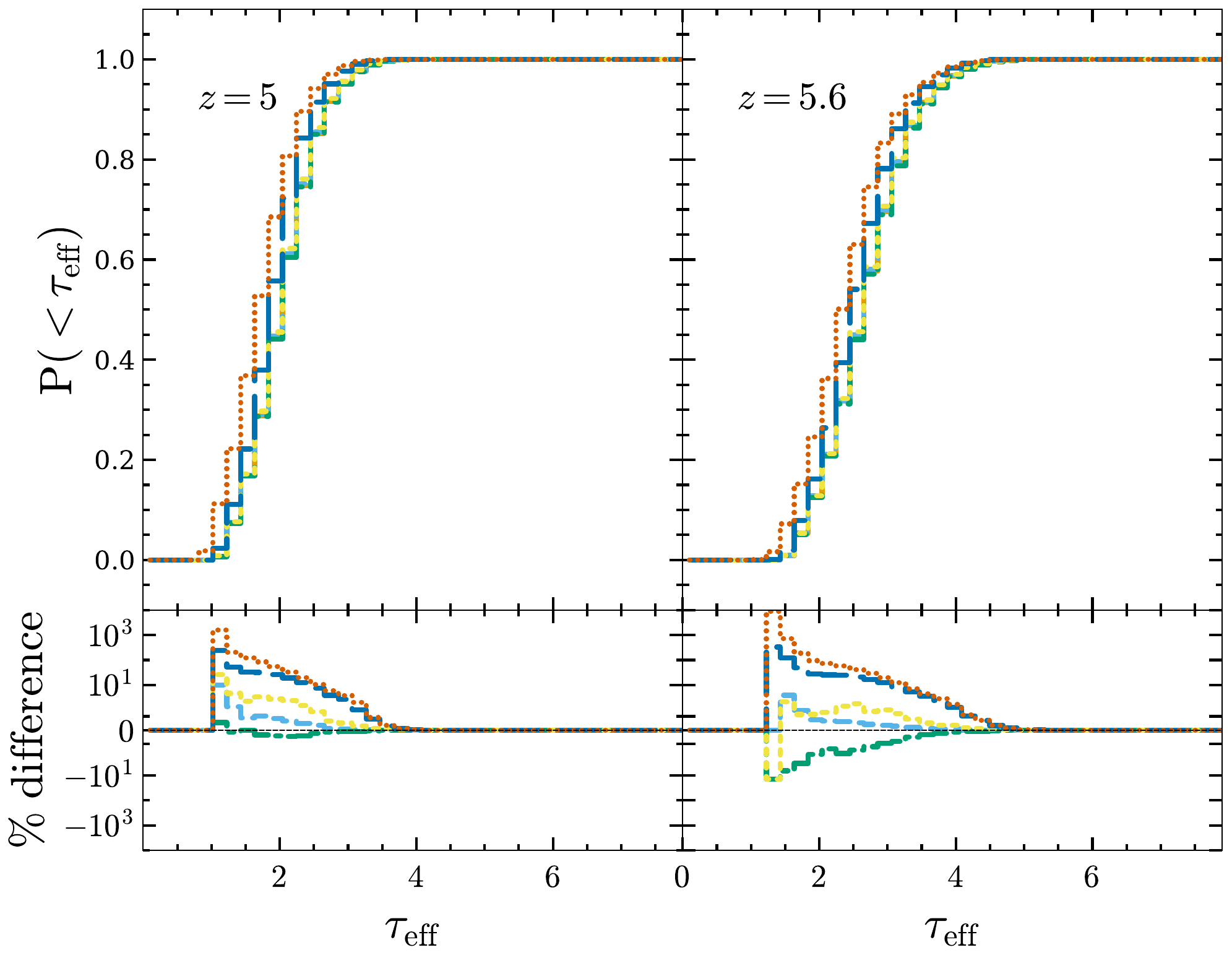}
    \caption{{\it Top panels:} Cumulative distribution function of Ly$\alpha$ effective optical depth as $z=5$ and 5.6 for \texttt{SS} (solid orange curve), \texttt{SS+ISM+XRBs} (dashed sky blue), \texttt{BS+ISM+XRBs} (dash-dotted green), \texttt{BB} (short-dashed yellow), \texttt{SOFT-PL} (long-dashed blue) and \texttt{AGN-PL} (dotted vermilion). {\it Bottom panels:} Fractional difference with respect to the \texttt{SS} simulation. }
    \label{fig:taueff_cdf}
\end{figure}

\subsubsection{Effective optical depth}

The first statistical quantity we discuss is the Ly$\alpha$ effective optical depth ($\tau_{\rm eff}$), a widely used characterization of the Ly$\alpha$ forest. To calculate this, we combine multiple synthetic spectra to obtain a length of 50$\rm{\mathit{h}^{-1} cMpc}$, traditionally adopted to represent observational data (e.g. \citealt{Bosman2022}), resulting in over $\sim 5500$ sightlines at each redshift. The effective optical depth is then computed as:
\begin{equation}
\tau_{\rm{eff}} = - \ln (\langle F \rangle),
\label{columndensity}
\end{equation}
where $\langle F \rangle$ is the mean transmitted flux of all the pixels in each new sightline. Figure \ref{fig:taueff_evol} displays the redshift evolution of the median $\tau_{\rm eff}$, along with the central $95 \%$ of the data, as well as a compilation of observational constraints from \citealt{Becker2015}, \citealt{eilers2018}, \citealt{bosman2018}, \citealt{yang2020} and \citealt{Bosman2022}. It is important to note that the input ionizing photon production rate in the \texttt{SS} simulation is calibrated to match the observational estimates of $\langle \rm F \rangle$ at $z \lesssim 5.5$ \citep{Becker2015,eilers2018,bosman2018,Bosman2022}. As the ionizing photon production rate is the same in all simulations, this ensures that the reference model is consistent with empirical data, while allowing us to examine deviations in the other models. All simulations produce a very similar behaviour, with $\tau_{\rm eff}$ decreasing with the progress of reionization, and slightly flattening at $z\lesssim6$. It is also interesting to see that all models are consistent with the majority of observational data within the scatter.

As can be better appreciated in the bottom panel of the figure, for $z \gtrsim 6$, the differences between models, with the exception of {\tt AGN-PL} and {\tt SOFT-PL}, are modest, with a maximum deviation from the {\tt SS} model of $\sim 1\%$, which becomes $<0.5\%$ at $z\sim5$ for {\tt SS+ISM+XRBs} and {\tt BS+ISM+XRBs}. This suggests that the SEDs of BS, ISM and XRBs do not significantly alter the effective optical depth. The deviations remain at the 1\% level also for a black-body spectrum. 
On the other hand, both {\tt SOFT-PL} and {\tt AGN-PL} at $z=5$ produce an effective optical depth  $\sim 10\%$ lower than the {\tt SS} one, highlighting the impact of the larger abundance of more energetic photons, which raise the  gas temperature, reducing the neutral fraction because of the smaller recombination rate, thus resulting in a higher transmission of Ly$\alpha$ photons in comparison to the other models. 

To further examine the differences among the SED models in terms of $\tau_{\rm eff}$, Figure \ref{fig:taueff_cdf} presents the cumulative distribution functions (CDFs) at three redshifts: $z = 6$, $5.5$, and $5$. We emphasize that our simulated optical depths are not calibrated to reproduce any observed CDF or mean transmission flux. As expected, the CDF systematically shifts toward lower optical depths as reionization progresses and more regions become ionized. When comparing different SED models to the \texttt{SS} reference, we find negligible deviations at both the high-$\tau_{\rm eff}$ end ($\tau_{\rm eff} > 6, 5, 4$ for $z = 6, 5.5,$ and $5$, respectively, where the cumulative probability is already unity) and the low-$\tau_{\rm eff}$ end, where the cumulative probability remains near zero. The differences are most apparent at intermediate $\tau_{\rm eff}$ values. Most models deviate from \texttt{SS} by less than $5\%$, whereas \texttt{SOFT-PL} and \texttt{AGN-PL} exhibit much larger discrepancies, reaching up to $\sim 1000\%$, with the latter showing the strongest effect.

Overall, aside from \texttt{SOFT-PL} and \texttt{AGN-PL}, all other models yield very similar Ly$\alpha$ forest effective optical depth statistics, particularly in terms of the median and scatter of $\tau_{\rm eff}$. Nonetheless, it remains interesting to investigate how these models influence small- and large-scale fluctuations, which we address in the next section.

\subsubsection{1-D flux power spectra}
\label{lya_ps}

\begin{figure}   
    \includegraphics[width=1.05\columnwidth]{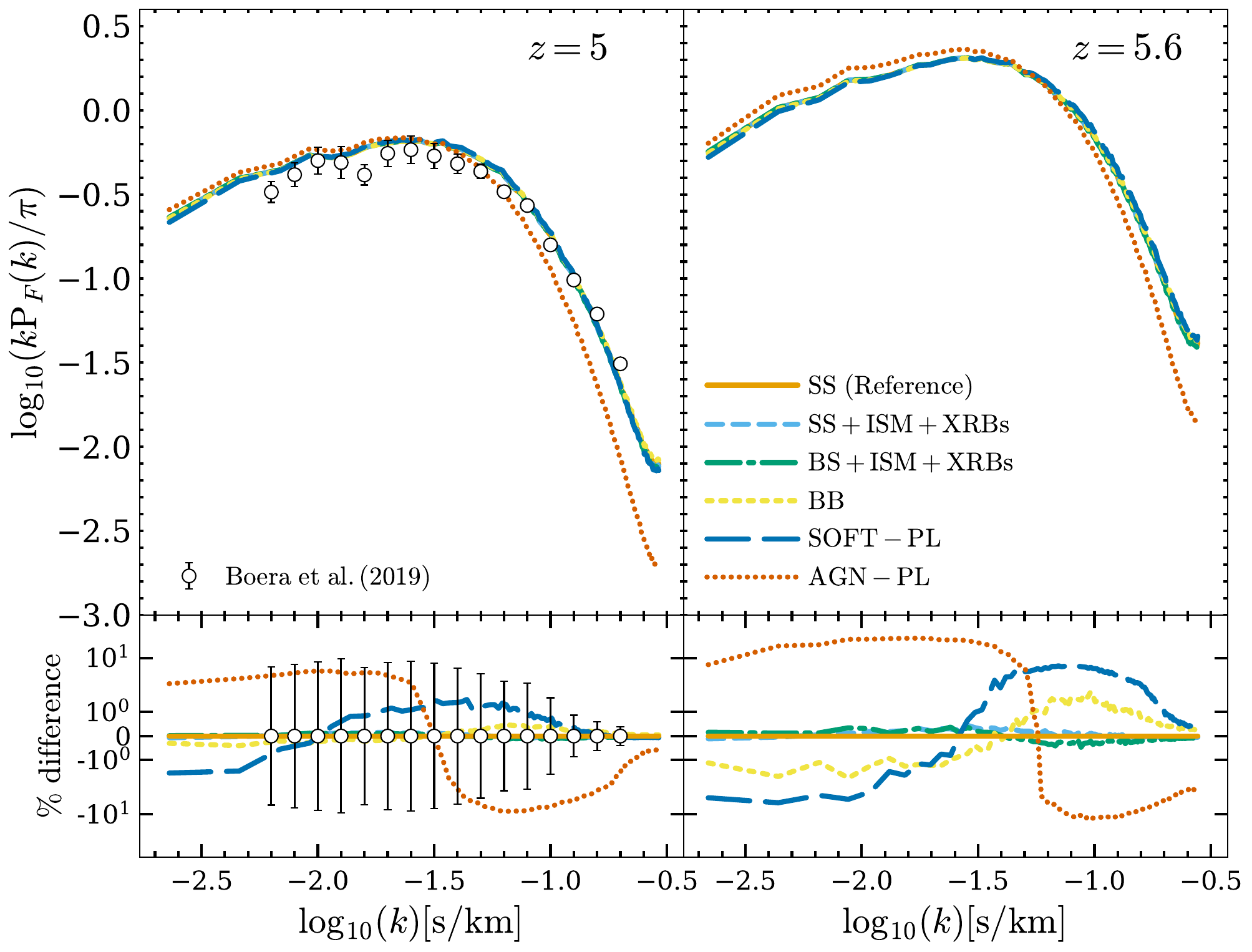}
    \caption{{\it Top panels:} 1-D dimensionless Ly$\alpha$ forest transmission flux power spectra at $z=5$ (\textit{left panel}) and 5.6 (\textit{right}) for \texttt{SS} (solid orange curve), \texttt{SS+ISM+XRBs} (dashed sky blue), \texttt{BS+ISM+XRBs} (dash-dotted green), \texttt{BB} (short-dashed yellow), \texttt{SOFT-PL} (long-dashed blue) and \texttt{AGN-PL} (dotted vermilion). The symbols denote a compilation of observational constraints from the literature \citep{boera2019}.
    {\it Bottom panel:} Fractional difference with respect to  the \texttt{SS} simulation along with the uncertainties of the observational constraints.}
   \label{fig:ps_w_scaling}
\end{figure}

Here we calculate the 1-D power spectrum (PS) of Ly$\alpha$ transmission flux using all sightlines from our simulation, and plot it in Figure~\ref{fig:ps_w_scaling} at $z=5$ and $z=5.6$. The latter redshift is chosen because it is when the largest differences  in $\langle \rm{\mathit{x}_{HI} \rangle}$ emerge, and at the same time there are enough regions of high transmission to render our exploration of their fluctuations more statistically significant. Note that here we have rescaled the mean transmitted flux in all simulations so that the redshift evolution of the mean transmission matches the observational constraints of \citet{Bosman2022}.  The overall shape of the power spectra is similar at both redshifts, increasing from $k \sim 10^{-0.5} \ \mathrm{s \ km^{-1}}$ to $k \sim 10^{-1.6} \ \mathrm{s \ km^{-1}}$, which marks the typical scale of fluctuations in the Ly$\alpha$ forest, and then decreasing at larger scales. However, the power at $z=5$ is lower than at $z=5.6$ because by then the IGM is mostly ionized, making the Ly$\alpha$ transmission more uniform. 
The suppression of power at the highest $k$-values comes mostly from  the thermal broadening kernel.  

At \( z = 5.6 \), models incorporating more energetic sources, i.e. \texttt{SS+ISM+XRBs} and \texttt{BS+ISM+XRBs}, exhibit only a slight deviation from the baseline \texttt{SS} model, showing a modest \( 0.5\% \) increase in the flux power spectrum at \( k \sim 10^{-1.6} \, \mathrm{s \, km^{-1}} \). Other SED models, such as \texttt{BB} and \texttt{SOFT-PL}, display more pronounced differences. These two models share a similar trend: an enhancement in power at small spatial scales and a suppression at larger scales compared to the \texttt{SS} model, with deviations reaching up to \( 8\% \) in the case of \texttt{SOFT-PL}. The decrease in power at small scales arises from the combined effects of ionization, recombination, and photoheating in dense regions, which amplify fluctuations in those environments. In contrast, at larger scales the partial ionization and associated photoheating lead to reduced fluctuations in the power spectrum for the \texttt{BB} and \texttt{SOFT-PL} models. 
Interestingly, the \texttt{AGN-PL} model stands out as at low $k$ it has about $10\%$ higher fluctuations than \texttt{SS} due to a more pronounced impact from the inhomogeneous reionization and heating process (see also, \citealt{Molaro2022}), while at higher $k$, the deviation in power drops before peaking again in the negative side at a $10\%$ deviation, driven by additional heating from helium double ionization. As the UV background becomes more homogeneous by $z=5$, fluctuations diminish, and differences in comparison to \texttt{SS} reach a maximum of $2\%$, mainly influenced by more pronounced double ionization of helium affecting the thermal Doppler broadening because of the increased IGM temperatures. By this stage, the deviations for \texttt{AGN-PL} remains at a higher value of $\sim 10\%$.

We note that all models are consistent with observational data from \citet{boera2019} at $z=5$, with the exception of \texttt{AGN-PL} at $k \gtrsim 10^{-1.2} \mathrm{s \, km^{-1}}$. Such small differences with respect to our reference \texttt{SS} model remain within the observational uncertainties at $z=5$, with the exception of the extreme \texttt{AGN-PL} model.
Here it is important to note that our Ly$\alpha$ forest analysis is based on post-processed radiative transfer, so that variations in the gas temperature between models do not feed back onto the gas density field. As a result, pressure (Jeans) smoothing driven by differences in the integrated thermal history is not self-consistently captured, and our results include only the effects of thermal Doppler broadening. Previous hydrodynamical studies that explicitly model pressure smoothing (e.g.\ \citealt{Molaro2022}) indicate that plausible variations in the thermal history can alter the Ly$\alpha$ 1-D flux power spectrum by up to $\sim$5--10\% at small scales. Given that the SED-driven differences reported here are of comparable or smaller magnitude, the scale dependence differs slightly. We therefore expect pressure smoothing to modestly rescale the amplitude of the small-scale differences in the power spectrum, but not to alter the qualitative trends between models. A fully self-consistent treatment would require coupled radiation--hydrodynamical simulations, which we leave for future work.

While current observations make it difficult to isolate the impact of different SEDs on the flux power spectrum, surveys with next-generation facilities like the Extremely Large Telescope (\texttt{ELT}, see \citealt{Dodorico2024}) and upgraded Dark Energy Spectroscopic Instrument (\texttt{DESI}, see \citealt{Karacayli2020}) could reduce the error bars to $1-2\%$, enabling a more precise assessment of how to distinguish the contribution from sources with different SEDs, and help in understanding if the assumptions on the sources SED in the theoretical models used to interpret Ly$\alpha$ forest data might induce any bias.

\subsubsection{Source-IGM connection}
\label{gal-igm}

\begin{figure*}  
\centering
\includegraphics[width=170mm]{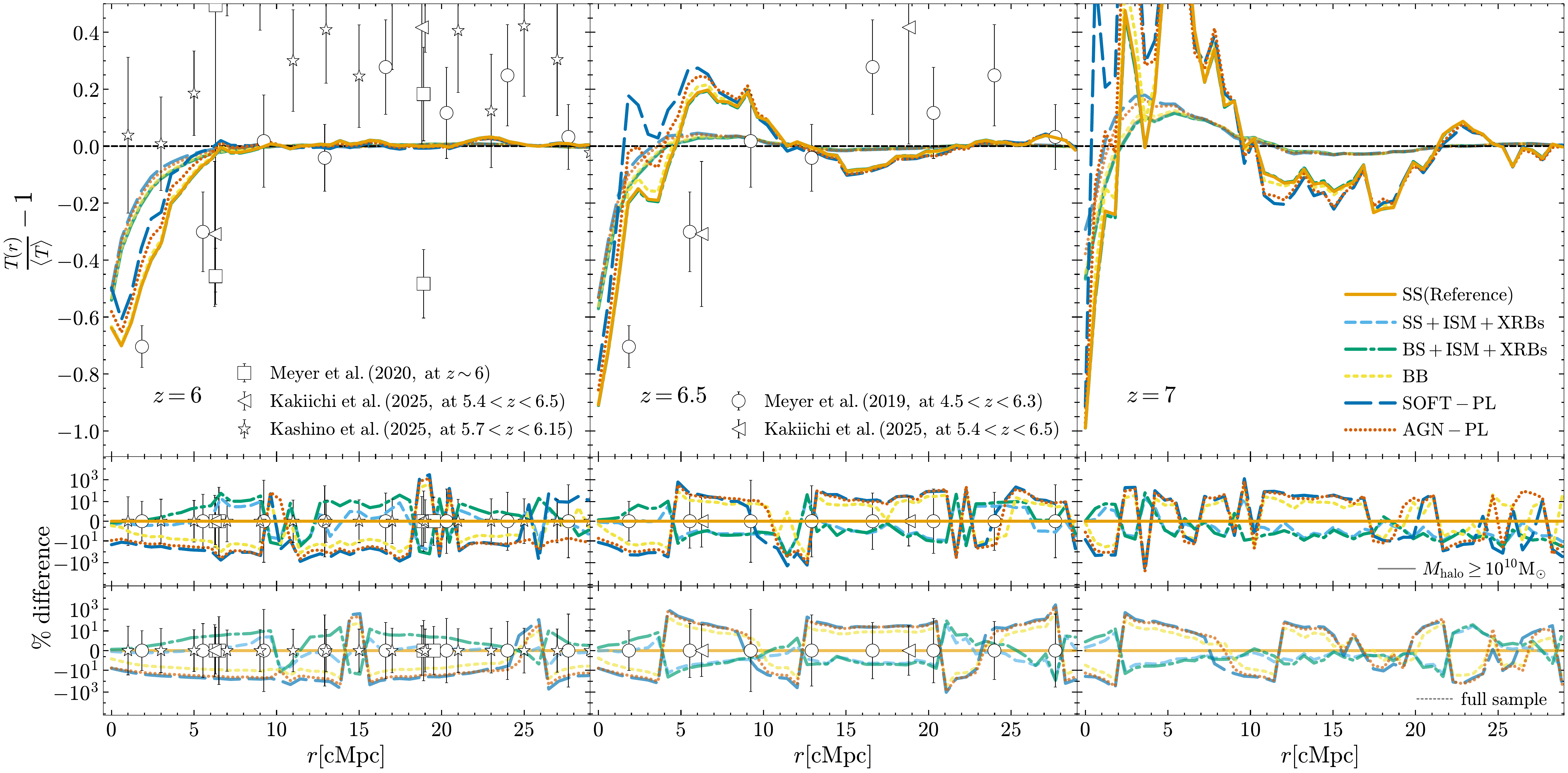}
    \caption{{\it Top panels:} Excess Ly$\alpha$ forest transmission at $z=6$ (left panel), 6.5 (middle) and 7 (right) for \texttt{SS} (solid orange curve), \texttt{SS+ISM+XRBs} (dashed sky blue), \texttt{BS+ISM+XRBs} (dash-dotted green), \texttt{BB} (short-dashed yellow), \texttt{SOFT-PL} (long-dashed blue) and \texttt{AGN-PL} (dotted vermilion). 
    The darker (lighter) curves represent all haloes (only haloes with $\rm{\mathit{M}_{halo} \geq 10^{10} M_{\odot}}$) The symbols denote a compilation of observational constraints from the literature \citep{meyer2019,meyer2020,koki2025,Kashino2025}. 
    {\it Bottom panel:} Fractional difference with respect to  \texttt{SS} simulation along with the uncertainties of the observational constraints for the sample of haloes with $\rm{\mathit{M}_{halo} \geq 10^{10} M_{\odot}}$ (upper part) and full sample of haloes (lower part).}
    \label{fig:excess_transmission}
\end{figure*}

To investigate more in detail the local effect of different SED models, 
we examine how the average IGM transmissivity ($T$) (i.e. the ratio between the transmitted flux and the inferred continuum flux at the same location) varies with distance $r$ from nearby sources. This approach, first proposed by \citet{kakiichi2018} and subsequently adopted also by \citet{meyer2019}, \citet{meyer2020} \citet{kashino2023} and \citet{koki2025}, provides a complementary way to probe the relation between sources and their surrounding IGM.
A key finding from these works is that the Ly$\alpha$ transmitted flux shows a broad peak at intermediate distances from ionizing sources (i.e. 10 $\rm{cMpc}/h$ $\lesssim$ $r$ $\lesssim$ 30 $\rm{cMpc}/h$), which can be interpreted as due to a proximity effect, suggesting that galaxies enhance ionization in their immediate surroundings, leading to increased Ly$\alpha$ transmissivity. While observational studies have made significant progress in this area \citep{koki2025,Kashino2025}, theoretical investigations remain limited, with the notable exception of \citet{garaldi2022} and \citet{garaldi2024}, who examined this phenomenon using high-resolution radiation-hydrodynamics simulations. More recently, \citet{conaboy2025} examined this connection between high-redshift galaxies and Ly$\alpha$ transmission utilising models of \texttt{Sherwood-relics} simulation.

Following the efforts made by \citet{garaldi2022} to create a synthetic version of observations, in Figure \ref{fig:excess_transmission} we present the excess IGM transmissivity relative to its average across all $r$, expressed as $ T(r) / \langle T \rangle - 1 $, at $ z = 6 $, 6.5, and 7. Note that, although the hydrodynamical simulation volume is periodic, the radiative transfer is performed in post-processing without imposing periodic boundary conditions as we discussed in Section \ref{section:2.3}. Hence, the maximum distance reaches upto $\sim30$ cMpc which corresponds to the simulation boxsize. The general trend of the curves is mostly the same at all redshifts. In the vicinity of the sources  they show a negative value (almost reaching -1, i.e. complete suppression of incoming flux), expressing an anti-correlation between source proximity and Ly$\alpha$ transmission. At larger radii, instead, they reach a peak indicating positive correlation, and finally, at $r \geq 20 \ \rm{cMpc}$, all models converge to zero since the local transmission becomes the same as the average one. 
Our simulations reveal an excess transmissivity at $ r \leq 10 $ cMpc. As redshift increases, we observe that the amplitude of the excess transmission increases and shifts to larger distances. This behaviour arises due to three key factors, namely that  at earlier times there are: (i) lower overdensities, which limit the suppression of Ly$\alpha$ flux near sources due to weaker hydrogen recombination; (ii) smaller ionized bubbles, which restrict the distance at which flux is enhanced; and (iii) a rapid decline in the average transmissivity $ \langle T \rangle $, while highly ionized regions continue to produce localized transmission spikes. These factors collectively shape the evolution of Ly$\alpha$ transmissivity and its dependence on source proximity over cosmic time. 

In the bottom panels of Figure \ref{fig:excess_transmission}, we show the fractional difference in excess transmission relative to the \texttt{SS} model. At $z = 6$, in the regions closest to the sources ($r \lesssim 9~\rm{cMpc}$), the \texttt{AGN-PL}, \texttt{SOFT-PL}, and \texttt{BB} models exhibit the strongest excess transmission, peaking at small distances before gradually declining outward. In contrast, the \texttt{BS+ISM+XRBs} model yields the lowest excess transmission, while the \texttt{SS+ISM+XRBs} model transitions from negative deviations at $ r \lesssim 6.5~\rm{cMpc}$ to positive deviations at larger radii. At $z = 6.5$, the transition to positive excess transmission shifts to $r \lesssim 4~\rm{cMpc}$, and by $z = 7$  to $r \leq 2~\rm{cMpc}$. In the regions with positive excess transmission, the \texttt{AGN-PL}, \texttt{SOFT-PL}, and \texttt{BB} models consistently produce the largest deviations, showing a peak at intermediate distances that is most prominent at $z = 6.5$ and $7$, while the overall excess transmission decreases by $z = 6$. Meanwhile, the \texttt{BS+ISM+XRBs} and \texttt{SS+ISM+XRBs} models remain slightly negative, with deviations $\sim 10\%$ lower than \texttt{SS}.

To provide a more meaningful comparison to observations, which are typically biased towards the brightest and most massive galaxies, we also repeat this calculation for a subset of haloes with $\rm{\mathit{M}_{halo}\geq 10^{10} M_{\odot}}$. The trends across SED models remain broadly similar to the full halo sample, but the amplitude of the excess transmissivity is higher, and the curves are noticeably noisier due to the small number of haloes in this mass range, particularly at $z=7$.

In the same figure, we also show the available observational measurements of excess Ly$\alpha$ transmission. These include the C~{\sc iv}-selected galaxies from \citet{meyer2019}, the Ly$\alpha$ emitters (LAEs) and Lyman-break galaxies (LBGs) from \citet{meyer2020}, the [O~{\sc iii}] emitters from \citet{Kashino2025}, and the quasar–galaxy cross-correlation measurements from \citet{koki2025}. The different galaxy selection methods naturally probe distinct environments, which contributes to the diversity in their measured profiles. 
At $z = 6$, our models exhibit a stronger deficit of Ly$\alpha$ transmission (i.e. negative value) at small radii than the measurements of \citet{Kashino2025}, but a weaker deficit than reported by \citet{meyer2019}, \citet{meyer2020} and \citet{koki2025}. The simulated profiles converge at $r\sim10$ cMpc, where the observations start showing the excess Ly$\alpha$ transmission. Beyond that, the excess Ly$\alpha$ transmission in the simulated models is marginal, differently from  observations, and only at $r\gtrsim20$ cMpc we observe a slightly better consistency.
At $z = 6.5$, our simulations show excess transmission at distances smaller than the \citet{meyer2019} and \citet{koki2025} measurements, while are more consistent with the shape reported by them. Again, at $r \gtrsim 20 \ \rm{cMpc}$ our predictions are in broad agreement with observational points of \citet{meyer2019}, while \citet{koki2025} have a much higher excess transmission.
Nevertheless, our limited simulation volume prevents us from fully capturing the large-scale fluctuations probed by these observations (as $T(r)$ is forced to converge to the mean value on scales comparable to the box size). Therefore, we mostly focus on the relative differences between the models and compare with the observational uncertainties.
As the observational error bars are generally larger than the model-to-model deviations, this limits the ability to distinguish between SED scenarios. However, in some cases, most notably for the \texttt{AGN-PL}, \texttt{SOFT-PL}, and \texttt{BB} models at $z = 6$ and $6.5$, the predicted deviations exceed the observational uncertainties at intermediate distances ($r \sim 5$–15 $\rm{cMpc}$), suggesting that these models could be distinguishable with current or near-future data. Even at the largest radii, where the deviations decline, some instances remain comparable to observational error bars, highlighting that the excess Ly$\alpha$ transmissivity around galaxies can provide a promising test of different ionizing source models as more precise measurements become available \footnote{Note that, as the maximum distance reached corresponds to the size of the simulation volume, we do not have enough statistical sample of pairs at the highest radii.}. Although different models in our simulation have nearly identical reionization histories, these differences stem purely from the source properties and are not affected by the different timing of the reionization process.

\section{Conclusions and Discussion}
\label{discussion}
This work investigates the influence of various sources SED modelling, including X-ray binaries, Bremsstrahlung emission from shock-heated ISM, and binary stars, on the IGM properties during the final stages of the epoch of reionization, as probed by the Ly$\alpha$ forest. We perform a comparative study including also more idealized, commonly adopted, SEDs, such as power-law and  blackbody spectra. Using simulations obtained by post-processing outputs of a \texttt{Sherwood}-type hydrodynamic simulation \citep{Bolton2017} with the 3D radiative transfer code \texttt{CRASH} \citep{ciardi2001,maselli2003,maselli2009,graziani2013,hariharan2017,glatzle2019,glatzle2022}, we explored the impact of different SEDs on the ionization and thermal state of the IGM, as well as their   imprint on Ly$\alpha$ forest observables. Our major findings are:
\begin{itemize}
    \item While the broad morphology of the fully ionized H\,{\sc ii} regions is similar across all SED models, the details of the ionization and thermal structure differ. Models including harder spectral components (e.g. ISM, XRBs, AGN-like SEDs) produce more extended partially ionized regions, enhanced He~{\sc iii} fractions, and higher surrounding IGM temperatures, compared to the models dominated by stellar emission (single or binary stars, or blackbody-like spectra).

    \item The evolution of the volume-averaged neutral hydrogen fraction is very similar across all SED models. Small differences arise at $z \gtrsim 6$, where models including harder spectral components (e.g. ISM, XRBs) show a slightly lower $\langle x_{\rm HI} \rangle$ due to enhanced partial ionization and heating, while power-law models remain marginally more neutral. By the end of reionization ($z$ $\lesssim 5.5$), the \texttt{AGN-PL} and \texttt{SOFT-PL} models predict a universe more ionized than the other models.
    
    \item The IGM temperature at mean density increases as reionization progresses, peaking around $z$ $\approx 6$ for most models. The \texttt{SS+ISM+XRBs} and \texttt{BS+ISM+XRBs} configurations show a modest increase in temperature compared to \texttt{SS}, driven by additional photoheating. Differently, the \texttt{BB} model results in slightly cooler temperatures. The \texttt{SOFT-PL} model has a behaviour similar to the \texttt{BB} one, but towards the end of reionization the gas becomes hotter than all the previous models because of  helium double ionization heating. The \texttt{AGN-PL} model consistently produces much higher temperatures, exceeding \texttt{SS} by  more than 7500 K at $z$ = 5, due to the large fraction of high-energy photons in its SED which are very efficient in fully ionizing helium. This scenario, though, is inconsistent with observational constraints.

    \item All models reproduce the general decline with redshift of the Ly$\alpha$ effective optical depth ($\tau_{\rm eff}$) obtained from observational data. Differences among models at  $z$ $\lesssim$ 6 are generally within 1$\%$, with the exception of \texttt{AGN-PL} and \texttt{SOFT-PL} which produce $\tau_{\rm eff}$ values up to 10$\%$ lower than those of the reference \texttt{SS} model due to their higher ionization and heating efficiency.
    
    \item The 1D power spectrum of Ly$\alpha$ transmission flux exhibits differences similar to the other quantities, with most models showing deviations of less than 2$\%$ from \texttt{SS}, mainly localized at intermediate scales, due to thermal broadening. Differently, \texttt{AGN-PL} produces up to 10$\%$ higher power on large scales due to spatial ionization variations, and 10$\%$ lower power on small scales from stronger heating. While subtle, these signatures could be observable with future high-precision spectroscopic surveys (i.e. \texttt{ELT} or \texttt{DESI}).

    \item By analyzing Ly$\alpha$ transmission as a function of distance from sources, we observe enhanced transmissivity (proximity effect) within $\sim$10 cMpc of massive halos. This effect is strongest at earlier times ($z=$ 7), when ionized regions are smaller and the IGM is more neutral. Hard-spectrum sources like \texttt{AGN-PL} and \texttt{SOFT-PL} produce the strongest proximity effects, while stellar models, particularly BS+ISM+XRBs, show more modest enhancements. These results are broadly consistent with recent observational studies (in particular at lower $r<10$ cMpc and higher radius $r>20$ cMpc), and suggest that proximity zone statistics could distinguish between different source populations in the reionization era.
\end{itemize}

Our findings emphasize that the choice of SED modeling is critical for interpreting Ly$\alpha$ forest measurements during the late stages of reionization. Adopting simplified spectra can bias the inferred impact of high energy sources, mis-estimating their role in shaping the thermal and ionization history of the IGM, which in turn can potentially affect the derivation and interpretation of Ly$\alpha$ forest observable quantities. However, the differences across models, while often modest in global statistics, are most pronounced in proximity zone behavior and the small- to intermediate-scale flux power spectrum, which represent the most promising observables to disentangle source populations. With the advent of next-generation spectroscopic surveys such as \texttt{ELT} and future \texttt{DESI} campaigns, improved measurements of these statistics will allow us to test the contribution of X-ray binaries, shock-heated ISM, and other high-energy sources. Utilising such data with physically motivated SED models will be essential to robustly connect Ly$\alpha$ forest observations to the sources driving the end of reionization and to refine our understanding of the IGM’s thermal evolution.

\section*{Acknowledgements}
All simulations were carried out on the machines of Max Planck Institute for Astrophysics (MPA) and Max Planck Computing and Data Facility (MPCDF). We thank the anonymous reviewer and the editor for the useful comments which helped to improve the manuscript. AB thanks the entire EoR research group of MPA for all the encouraging comments for this project. We thank
Daichi Kashino for sharing the \texttt{EIGERS-VII} data and Luke Conaboy for sharing the compilation of the data for the galaxy-IGM connection. This work made extensive use of publicly available software packages : \texttt{numpy} \citep{vander2011}, \texttt{matplotlib} \citep{Hunter2007}, \texttt{scipy} \citep{Jones2001} and \texttt{CoReCon} \citep{Garaldi2023}. Authors thank the developers of these packages. JSB is supported by STFC consolidated grant ST/X000982/1.
MV is supported by the INFN INDARK and INAF Theory "Cosmological Investigation of the  cosmic web" grants. 
EG is supported by the JSPS KAKENHI grant ILR 23K20035. 
\label{acknowledgement}

\section*{Data Availability}

The final data products from this study will be shared on reasonable request to the corresponding author.




\bibliographystyle{mnras}
\bibliography{mnras} 

\begin{thebibliography}{}
\makeatletter
\relax
\def\mn@urlcharsother{\let\do\@makeother \do\$\do\&\do\#\do\^\do\_\do\%\do\~}
\def\mn@doi{\begingroup\mn@urlcharsother \@ifnextchar [ {\mn@doi@} {\mn@doi@[]}}
\def\mn@doi@[#1]#2{\def\@tempa{#1}\ifx\@tempa\@empty \href {http://dx.doi.org/#2} {doi:#2}\else \href {http://dx.doi.org/#2} {#1}\fi \endgroup}
\def\mn@eprint#1#2{\mn@eprint@#1:#2::\@nil}
\def\mn@eprint@arXiv#1{\href {http://arxiv.org/abs/#1} {{\tt arXiv:#1}}}
\def\mn@eprint@dblp#1{\href {http://dblp.uni-trier.de/rec/bibtex/#1.xml} {dblp:#1}}
\def\mn@eprint@#1:#2:#3:#4\@nil{\def\@tempa {#1}\def\@tempb {#2}\def\@tempc {#3}\ifx \@tempc \@empty \let \@tempc \@tempb \let \@tempb \@tempa \fi \ifx \@tempb \@empty \def\@tempb {arXiv}\fi \@ifundefined {mn@eprint@\@tempb}{\@tempb:\@tempc}{\expandafter \expandafter \csname mn@eprint@\@tempb\endcsname \expandafter{\@tempc}}}

\bibitem[\protect\citeauthoryear{{Akiyama} et~al.,}{{Akiyama} et~al.}{2018}]{Akiyama2018}
{Akiyama} M.,  et~al., 2018, \mn@doi [\pasj] {10.1093/pasj/psx091}, \href {https://ui.adsabs.harvard.edu/abs/2018PASJ...70S..34A} {70, S34}

\bibitem[\protect\citeauthoryear{{Asthana}, {Haehnelt}, {Kulkarni}, {Bolton}, {Gaikwad}, {Keating}  \& {Puchwein}}{{Asthana} et~al.}{2024}]{Asthana2024}
{Asthana} S.,  {Haehnelt} M.~G.,  {Kulkarni} G.,  {Bolton} J.~S.,  {Gaikwad} P.,  {Keating} L.~C.,   {Puchwein} E.,  2024, \mn@doi [arXiv e-prints] {10.48550/arXiv.2409.15453}, \href {https://ui.adsabs.harvard.edu/abs/2024arXiv240915453A} {p. arXiv:2409.15453}

\bibitem[\protect\citeauthoryear{Baek, Semelin, Di~Matteo, Revaz  \& Combes}{Baek et~al.}{2010}]{Baek2010}
Baek S.,  Semelin B.,  Di~Matteo P.,  Revaz Y.,   Combes F.,  2010, \mn@doi [Astronomy & Astrophysics] {10.1051/0004-6361/201014347}, 523, A4

\bibitem[\protect\citeauthoryear{{Barnes}, {Wood}, {Hill}  \& {Haffner}}{{Barnes} et~al.}{2014}]{Barnes2014}
{Barnes} J.~E.,  {Wood} K.,  {Hill} A.~S.,   {Haffner} L.~M.,  2014, \mn@doi [\mnras] {10.1093/mnras/stu521}, \href {https://ui.adsabs.harvard.edu/abs/2014MNRAS.440.3027B} {440, 3027}

\bibitem[\protect\citeauthoryear{{Basu}, {Garaldi}  \& {Ciardi}}{{Basu} et~al.}{2024}]{Basu2024}
{Basu} A.,  {Garaldi} E.,   {Ciardi} B.,  2024, \mn@doi [\mnras] {10.1093/mnras/stae1488}, \href {https://ui.adsabs.harvard.edu/abs/2024MNRAS.532..841B} {532, 841}

\bibitem[\protect\citeauthoryear{{Becker} \& {Bolton}}{{Becker} \& {Bolton}}{2013}]{BeckerBolton2013}
{Becker} G.~D.,  {Bolton} J.~S.,  2013, \mn@doi [\mnras] {10.1093/mnras/stt1610}, \href {https://ui.adsabs.harvard.edu/abs/2013MNRAS.436.1023B} {436, 1023}

\bibitem[\protect\citeauthoryear{{Becker}, {Bolton}  \& {Lidz}}{{Becker} et~al.}{2015}]{Becker2015}
{Becker} G.~D.,  {Bolton} J.~S.,   {Lidz} A.,  2015, \mn@doi [\pasa] {10.1017/pasa.2015.45}, \href {https://ui.adsabs.harvard.edu/abs/2015PASA...32...45B} {32, e045}

\bibitem[\protect\citeauthoryear{{Becker}, {D'Aloisio}, {Christenson}, {Zhu}, {Worseck}  \& {Bolton}}{{Becker} et~al.}{2021}]{becker2021}
{Becker} G.~D.,  {D'Aloisio} A.,  {Christenson} H.~M.,  {Zhu} Y.,  {Worseck} G.,   {Bolton} J.~S.,  2021, \mn@doi [\mnras] {10.1093/mnras/stab2696}, \href {https://ui.adsabs.harvard.edu/abs/2021MNRAS.508.1853B} {508, 1853}

\bibitem[\protect\citeauthoryear{{Bera}, {Samui}  \& {Datta}}{{Bera} et~al.}{2023}]{bera2023}
{Bera} A.,  {Samui} S.,   {Datta} K.~K.,  2023, \mn@doi [\mnras] {10.1093/mnras/stac3814}, \href {https://ui.adsabs.harvard.edu/abs/2023MNRAS.519.4869B} {519, 4869}

\bibitem[\protect\citeauthoryear{{Bhagwat}, {Costa}, {Ciardi}, {Pakmor}  \& {Garaldi}}{{Bhagwat} et~al.}{2024}]{bhagwat2024}
{Bhagwat} A.,  {Costa} T.,  {Ciardi} B.,  {Pakmor} R.,   {Garaldi} E.,  2024, \mn@doi [\mnras] {10.1093/mnras/stae1125}, \href {https://ui.adsabs.harvard.edu/abs/2024MNRAS.531.3406B} {531, 3406}

\bibitem[\protect\citeauthoryear{{Boera}, {Becker}, {Bolton}  \& {Nasir}}{{Boera} et~al.}{2019}]{boera2019}
{Boera} E.,  {Becker} G.~D.,  {Bolton} J.~S.,   {Nasir} F.,  2019, \mn@doi [\apj] {10.3847/1538-4357/aafee4}, \href {https://ui.adsabs.harvard.edu/abs/2019ApJ...872..101B} {872, 101}

\bibitem[\protect\citeauthoryear{Bolton, Becker, Raskutti, Wyithe, Haehnelt  \& Sargent}{Bolton et~al.}{2012}]{Bolton2012}
Bolton J.~S.,  Becker G.~D.,  Raskutti S.,  Wyithe J. S.~B.,  Haehnelt M.~G.,   Sargent W. L.~W.,  2012, \mn@doi [Monthly Notices of the Royal Astronomical Society] {10.1111/j.1365-2966.2011.19929.x}, 419, 2880

\bibitem[\protect\citeauthoryear{{Bolton}, {Puchwein}, {Sijacki}, {Haehnelt}, {Kim}, {Meiksin}, {Regan}  \& {Viel}}{{Bolton} et~al.}{2017}]{Bolton2017}
{Bolton} J.~S.,  {Puchwein} E.,  {Sijacki} D.,  {Haehnelt} M.~G.,  {Kim} T.-S.,  {Meiksin} A.,  {Regan} J.~A.,   {Viel} M.,  2017, \mn@doi [\mnras] {10.1093/mnras/stw2397}, \href {https://ui.adsabs.harvard.edu/abs/2017MNRAS.464..897B} {464, 897}

\bibitem[\protect\citeauthoryear{Bosman, Fan, Jiang, Reed, Matsuoka, Becker  \& Haehnelt}{Bosman et~al.}{2018}]{bosman2018}
Bosman S. E.~I.,  Fan X.,  Jiang L.,  Reed S.,  Matsuoka Y.,  Becker G.,   Haehnelt M.,  2018, \mn@doi [Monthly Notices of the Royal Astronomical Society] {10.1093/mnras/sty1344}

\bibitem[\protect\citeauthoryear{{Bosman} et~al.,}{{Bosman} et~al.}{2022}]{Bosman2022}
{Bosman} S. E.~I.,  et~al., 2022, \mn@doi [\mnras] {10.1093/mnras/stac1046}, \href {https://ui.adsabs.harvard.edu/abs/2022MNRAS.514...55B} {514, 55}

\bibitem[\protect\citeauthoryear{{Boutsia}, {Grazian}, {Giallongo}, {Fiore}  \& {Civano}}{{Boutsia} et~al.}{2018}]{Boutsia2018}
{Boutsia} K.,  {Grazian} A.,  {Giallongo} E.,  {Fiore} F.,   {Civano} F.,  2018, \mn@doi [\apj] {10.3847/1538-4357/aae6c7}, \href {https://ui.adsabs.harvard.edu/abs/2018ApJ...869...20B} {869, 20}

\bibitem[\protect\citeauthoryear{Bouwens, Illingworth, Oesch, Caruana, Holwerda, Smit  \& Wilkins}{Bouwens et~al.}{2015}]{Bouwens2015}
Bouwens R.~J.,  Illingworth G.~D.,  Oesch P.~A.,  Caruana J.,  Holwerda B.,  Smit R.,   Wilkins S.,  2015, \mn@doi [The Astrophysical Journal] {10.1088/0004-637x/811/2/140}, 811, 140

\bibitem[\protect\citeauthoryear{Calverley, Becker, Haehnelt  \& Bolton}{Calverley et~al.}{2011}]{Calverley2011}
Calverley A.~P.,  Becker G.~D.,  Haehnelt M.~G.,   Bolton J.~S.,  2011, \mn@doi [Monthly Notices of the Royal Astronomical Society] {10.1111/j.1365-2966.2010.18072.x}, 412, 2543

\bibitem[\protect\citeauthoryear{{Cang}, {Gao}  \& {Ma}}{{Cang} et~al.}{2023}]{cang2023}
{Cang} J.,  {Gao} Y.,   {Ma} Y.-Z.,  2023, \mn@doi [arXiv e-prints] {10.48550/arXiv.2312.17499}, \href {https://ui.adsabs.harvard.edu/abs/2023arXiv231217499C} {p. arXiv:2312.17499}

\bibitem[\protect\citeauthoryear{{Carvalho}, {Escobar}  \& {Pellizza}}{{Carvalho} et~al.}{2024}]{carvalho2024}
{Carvalho} L.,  {Escobar} G.~J.,   {Pellizza} L.~J.,  2024, Boletin de la Asociacion Argentina de Astronomia La Plata Argentina, \href {https://ui.adsabs.harvard.edu/abs/2024BAAA...65..240C} {65, 240}

\bibitem[\protect\citeauthoryear{{Chang}, {Broderick}  \& {Pfrommer}}{{Chang} et~al.}{2012}]{chang2012}
{Chang} P.,  {Broderick} A.~E.,   {Pfrommer} C.,  2012, \mn@doi [\apj] {10.1088/0004-637X/752/1/23}, \href {https://ui.adsabs.harvard.edu/abs/2012ApJ...752...23C} {752, 23}

\bibitem[\protect\citeauthoryear{{Chen}, {Stark}, {Mason}, {Topping}, {Whitler}, {Tang}, {Endsley}  \& {Charlot}}{{Chen} et~al.}{2024}]{chen2024}
{Chen} Z.,  {Stark} D.~P.,  {Mason} C.,  {Topping} M.~W.,  {Whitler} L.,  {Tang} M.,  {Endsley} R.,   {Charlot} S.,  2024, \mn@doi [\mnras] {10.1093/mnras/stae455}, \href {https://ui.adsabs.harvard.edu/abs/2024MNRAS.528.7052C} {528, 7052}

\bibitem[\protect\citeauthoryear{Chevalier \& Clegg}{Chevalier \& Clegg}{1985}]{Chevalier1985}
Chevalier R.~A.,  Clegg A.~W.,  1985, \mn@doi [Nature] {10.1038/317044a0}, 317, 44

\bibitem[\protect\citeauthoryear{{Choudhury} \& {Paranjape}}{{Choudhury} \& {Paranjape}}{2022}]{choudhury2022}
{Choudhury} T.~R.,  {Paranjape} A.,  2022, {SCRIPT: Semi-numerical Code for ReIonization with PhoTon-conservation}, Astrophysics Source Code Library, record ascl:2204.013

\bibitem[\protect\citeauthoryear{{Ciardi}, {Ferrara}, {Marri}  \& {Raimondo}}{{Ciardi} et~al.}{2001}]{ciardi2001}
{Ciardi} B.,  {Ferrara} A.,  {Marri} S.,   {Raimondo} G.,  2001, \mn@doi [\mnras] {10.1046/j.1365-8711.2001.04316.x}, \href {https://ui.adsabs.harvard.edu/abs/2001MNRAS.324..381C} {324, 381}

\bibitem[\protect\citeauthoryear{{Ciardi}, {Bolton}, {Maselli}  \& {Graziani}}{{Ciardi} et~al.}{2012}]{Ciardi2012}
{Ciardi} B.,  {Bolton} J.~S.,  {Maselli} A.,   {Graziani} L.,  2012, \mn@doi [\mnras] {10.1111/j.1365-2966.2012.20902.x}, \href {https://ui.adsabs.harvard.edu/abs/2012MNRAS.423..558C} {423, 558}

\bibitem[\protect\citeauthoryear{{Conaboy}, {Bolton}, {Keating}, {Haehnelt}, {Kulkarni}  \& {Puchwein}}{{Conaboy} et~al.}{2025}]{conaboy2025}
{Conaboy} L.,  {Bolton} J.~S.,  {Keating} L.~C.,  {Haehnelt} M.~G.,  {Kulkarni} G.,   {Puchwein} E.,  2025, \mn@doi [arXiv e-prints] {10.48550/arXiv.2502.02983}, \href {https://ui.adsabs.harvard.edu/abs/2025arXiv250202983C} {p. arXiv:2502.02983}

\bibitem[\protect\citeauthoryear{{D'Aloisio}, {McQuinn}, {Davies}  \& {Furlanetto}}{{D'Aloisio} et~al.}{2018}]{daloisio2018}
{D'Aloisio} A.,  {McQuinn} M.,  {Davies} F.~B.,   {Furlanetto} S.~R.,  2018, \mn@doi [\mnras] {10.1093/mnras/stx2341}, \href {https://ui.adsabs.harvard.edu/abs/2018MNRAS.473..560D} {473, 560}

\bibitem[\protect\citeauthoryear{{D'Odorico} et~al.,}{{D'Odorico} et~al.}{2024}]{Dodorico2024}
{D'Odorico} V.,  et~al., 2024, \mn@doi [Experimental Astronomy] {10.1007/s10686-024-09967-3}, \href {https://ui.adsabs.harvard.edu/abs/2024ExA....58...21D} {58, 21}

\bibitem[\protect\citeauthoryear{Davies, Hennawi, Eilers  \& Lukić}{Davies et~al.}{2018}]{Davies2018}
Davies F.~B.,  Hennawi J.~F.,  Eilers A.-C.,   Lukić Z.,  2018, \mn@doi [The Astrophysical Journal] {10.3847/1538-4357/aaaf70}, 855, 106

\bibitem[\protect\citeauthoryear{{Davies} et~al.,}{{Davies} et~al.}{2024}]{davies2024}
{Davies} F.~B.,  et~al., 2024, \mn@doi [\apj] {10.3847/1538-4357/ad1d5d}, \href {https://ui.adsabs.harvard.edu/abs/2024ApJ...965..134D} {965, 134}

\bibitem[\protect\citeauthoryear{{Dayal} \& {Ferrara}}{{Dayal} \& {Ferrara}}{2018}]{Pratika2018}
{Dayal} P.,  {Ferrara} A.,  2018, \mn@doi [\physrep] {10.1016/j.physrep.2018.10.002}, \href {https://ui.adsabs.harvard.edu/abs/2018PhR...780....1D} {780, 1}

\bibitem[\protect\citeauthoryear{Dayal et~al.,}{Dayal et~al.}{2020}]{dayal2020}
Dayal P.,  et~al., 2020, \mn@doi [Monthly Notices of the Royal Astronomical Society] {10.1093/mnras/staa1138}, 495, 3065–3078

\bibitem[\protect\citeauthoryear{{Draine}}{{Draine}}{1978}]{Draine1978}
{Draine} B.~T.,  1978, \mn@doi [\apjs] {10.1086/190513}, \href {https://ui.adsabs.harvard.edu/abs/1978ApJS...36..595D} {36, 595}

\bibitem[\protect\citeauthoryear{Eide, Graziani, Ciardi, Feng, Kakiichi  \& Di~Matteo}{Eide et~al.}{2018}]{Marius2018}
Eide M.~B.,  Graziani L.,  Ciardi B.,  Feng Y.,  Kakiichi K.,   Di~Matteo T.,  2018, \mn@doi [Monthly Notices of the Royal Astronomical Society] {10.1093/mnras/sty272}, 476, 1174–1190

\bibitem[\protect\citeauthoryear{{Eide}, {Ciardi}, {Graziani}, {Busch}, {Feng}  \& {Di Matteo}}{{Eide} et~al.}{2020a}]{Marius2020}
{Eide} M.~B.,  {Ciardi} B.,  {Graziani} L.,  {Busch} P.,  {Feng} Y.,   {Di Matteo} T.,  2020a, \mn@doi [\mnras] {10.1093/mnras/staa2774}, \href {https://ui.adsabs.harvard.edu/abs/2020MNRAS.498.6083E} {498, 6083}

\bibitem[\protect\citeauthoryear{{Eide}, {Ciardi}, {Feng}  \& {Di Matteo}}{{Eide} et~al.}{2020b}]{eide2020b}
{Eide} M.~B.,  {Ciardi} B.,  {Feng} Y.,   {Di Matteo} T.,  2020b, \mn@doi [\mnras] {10.1093/mnras/staa3253}, \href {https://ui.adsabs.harvard.edu/abs/2020MNRAS.499.5978E} {499, 5978}

\bibitem[\protect\citeauthoryear{{Eilers}, {Davies}  \& {Hennawi}}{{Eilers} et~al.}{2018}]{eilers2018}
{Eilers} A.-C.,  {Davies} F.~B.,   {Hennawi} J.~F.,  2018, \mn@doi [\apj] {10.3847/1538-4357/aad4fd}, \href {https://ui.adsabs.harvard.edu/abs/2018ApJ...864...53E} {864, 53}

\bibitem[\protect\citeauthoryear{{Fan}, {Carilli}  \& {Keating}}{{Fan} et~al.}{2006}]{Fan2006}
{Fan} X.,  {Carilli} C.~L.,   {Keating} B.,  2006, \mn@doi [\araa] {10.1146/annurev.astro.44.051905.092514}, \href {https://ui.adsabs.harvard.edu/abs/2006ARA&A..44..415F} {44, 415}

\bibitem[\protect\citeauthoryear{{Finlator}, {Keating}, {Oppenheimer}, {Dav{\'e}}  \& {Zackrisson}}{{Finlator} et~al.}{2018}]{finlator2018}
{Finlator} K.,  {Keating} L.,  {Oppenheimer} B.~D.,  {Dav{\'e}} R.,   {Zackrisson} E.,  2018, \mn@doi [\mnras] {10.1093/mnras/sty1949}, \href {https://ui.adsabs.harvard.edu/abs/2018MNRAS.480.2628F} {480, 2628}

\bibitem[\protect\citeauthoryear{{Fletcher}, {Tang}, {Robertson}, {Nakajima}, {Ellis}, {Stark}  \& {Inoue}}{{Fletcher} et~al.}{2019}]{fletcher2019}
{Fletcher} T.~J.,  {Tang} M.,  {Robertson} B.~E.,  {Nakajima} K.,  {Ellis} R.~S.,  {Stark} D.~P.,   {Inoue} A.,  2019, \mn@doi [\apj] {10.3847/1538-4357/ab2045}, \href {https://ui.adsabs.harvard.edu/abs/2019ApJ...878...87F} {878, 87}

\bibitem[\protect\citeauthoryear{{Fragos} et~al.,}{{Fragos} et~al.}{2013a}]{fragos2013}
{Fragos} T.,  et~al., 2013a, \mn@doi [\apj] {10.1088/0004-637X/764/1/41}, \href {https://ui.adsabs.harvard.edu/abs/2013ApJ...764...41F} {764, 41}

\bibitem[\protect\citeauthoryear{{Fragos}, {Lehmer}, {Naoz}, {Zezas}  \& {Basu-Zych}}{{Fragos} et~al.}{2013b}]{fragosa2013}
{Fragos} T.,  {Lehmer} B.~D.,  {Naoz} S.,  {Zezas} A.,   {Basu-Zych} A.,  2013b, \mn@doi [\apjl] {10.1088/2041-8205/776/2/L31}, \href {https://ui.adsabs.harvard.edu/abs/2013ApJ...776L..31F} {776, L31}

\bibitem[\protect\citeauthoryear{{Gaikwad} et~al.,}{{Gaikwad} et~al.}{2020}]{Gaikwad2020}
{Gaikwad} P.,  et~al., 2020, \mn@doi [\mnras] {10.1093/mnras/staa907}, \href {https://ui.adsabs.harvard.edu/abs/2020MNRAS.494.5091G} {494, 5091}

\bibitem[\protect\citeauthoryear{{Gaikwad} et~al.,}{{Gaikwad} et~al.}{2023}]{gaikwad2023}
{Gaikwad} P.,  et~al., 2023, \mn@doi [\mnras] {10.1093/mnras/stad2566}, \href {https://ui.adsabs.harvard.edu/abs/2023MNRAS.525.4093G} {525, 4093}

\bibitem[\protect\citeauthoryear{{Garaldi}}{{Garaldi}}{2023}]{Garaldi2023}
{Garaldi} E.,  2023, \mn@doi [The Journal of Open Source Software] {10.21105/joss.05407}, \href {https://ui.adsabs.harvard.edu/abs/2023JOSS....8.5407G} {8, 5407}

\bibitem[\protect\citeauthoryear{{Garaldi} \& {Bellscheidt}}{{Garaldi} \& {Bellscheidt}}{2024}]{garaldi2024}
{Garaldi} E.,  {Bellscheidt} V.,  2024, \mn@doi [arXiv e-prints] {10.48550/arXiv.2410.02853}, \href {https://ui.adsabs.harvard.edu/abs/2024arXiv241002853G} {p. arXiv:2410.02853}

\bibitem[\protect\citeauthoryear{{Garaldi}, {Compostella}  \& {Porciani}}{{Garaldi} et~al.}{2019}]{enrico2019}
{Garaldi} E.,  {Compostella} M.,   {Porciani} C.,  2019, \mn@doi [\mnras] {10.1093/mnras/sty3414}, \href {https://ui.adsabs.harvard.edu/abs/2019MNRAS.483.5301G} {483, 5301}

\bibitem[\protect\citeauthoryear{{Garaldi}, {Kannan}, {Smith}, {Springel}, {Pakmor}, {Vogelsberger}  \& {Hernquist}}{{Garaldi} et~al.}{2022}]{garaldi2022}
{Garaldi} E.,  {Kannan} R.,  {Smith} A.,  {Springel} V.,  {Pakmor} R.,  {Vogelsberger} M.,   {Hernquist} L.,  2022, \mn@doi [\mnras] {10.1093/mnras/stac257}, \href {https://ui.adsabs.harvard.edu/abs/2022MNRAS.512.4909G} {512, 4909}

\bibitem[\protect\citeauthoryear{{Gessey-Jones}, {Fialkov}, {de Lera Acedo}, {Handley}  \& {Barkana}}{{Gessey-Jones} et~al.}{2023}]{gessey2023}
{Gessey-Jones} T.,  {Fialkov} A.,  {de Lera Acedo} E.,  {Handley} W.~J.,   {Barkana} R.,  2023, \mn@doi [\mnras] {10.1093/mnras/stad3014}, \href {https://ui.adsabs.harvard.edu/abs/2023MNRAS.526.4262G} {526, 4262}

\bibitem[\protect\citeauthoryear{{Giallongo} et~al.,}{{Giallongo} et~al.}{2015}]{Giallongo2015}
{Giallongo} E.,  et~al., 2015, \mn@doi [\aap] {10.1051/0004-6361/201425334}, \href {https://ui.adsabs.harvard.edu/abs/2015A&A...578A..83G} {578, A83}

\bibitem[\protect\citeauthoryear{{Giallongo} et~al.,}{{Giallongo} et~al.}{2019}]{Giallongo2019}
{Giallongo} E.,  et~al., 2019, \mn@doi [\apj] {10.3847/1538-4357/ab39e1}, \href {https://ui.adsabs.harvard.edu/abs/2019ApJ...884...19G} {884, 19}

\bibitem[\protect\citeauthoryear{{Glatzle}, {Ciardi}  \& {Graziani}}{{Glatzle} et~al.}{2019}]{glatzle2019}
{Glatzle} M.,  {Ciardi} B.,   {Graziani} L.,  2019, \mn@doi [\mnras] {10.1093/mnras/sty2514}, \href {https://ui.adsabs.harvard.edu/abs/2019MNRAS.482..321G} {482, 321}

\bibitem[\protect\citeauthoryear{{Glatzle}, {Graziani}  \& {Ciardi}}{{Glatzle} et~al.}{2022}]{glatzle2022}
{Glatzle} M.,  {Graziani} L.,   {Ciardi} B.,  2022, \mn@doi [\mnras] {10.1093/mnras/stab3459}, \href {https://ui.adsabs.harvard.edu/abs/2022MNRAS.510.1068G} {510, 1068}

\bibitem[\protect\citeauthoryear{{Gnedin}}{{Gnedin}}{2014}]{gnedin2014}
{Gnedin} N.~Y.,  2014, \mn@doi [\apj] {10.1088/0004-637X/793/1/29}, \href {https://ui.adsabs.harvard.edu/abs/2014ApJ...793...29G} {793, 29}

\bibitem[\protect\citeauthoryear{{Goulding} et~al.,}{{Goulding} et~al.}{2023}]{Goulding2023}
{Goulding} A.~D.,  et~al., 2023, \mn@doi [\apjl] {10.3847/2041-8213/acf7c5}, \href {https://ui.adsabs.harvard.edu/abs/2023ApJ...955L..24G} {955, L24}

\bibitem[\protect\citeauthoryear{{Grazian} et~al.,}{{Grazian} et~al.}{2024}]{grazian2024}
{Grazian} A.,  et~al., 2024, \mn@doi [\apj] {10.3847/1538-4357/ad6980}, \href {https://ui.adsabs.harvard.edu/abs/2024ApJ...974...84G} {974, 84}

\bibitem[\protect\citeauthoryear{{Graziani}, {Maselli}  \& {Ciardi}}{{Graziani} et~al.}{2013}]{graziani2013}
{Graziani} L.,  {Maselli} A.,   {Ciardi} B.,  2013, \mn@doi [\mnras] {10.1093/mnras/stt206}, \href {https://ui.adsabs.harvard.edu/abs/2013MNRAS.431..722G} {431, 722}

\bibitem[\protect\citeauthoryear{{Graziani}, {Ciardi}  \& {Glatzle}}{{Graziani} et~al.}{2018}]{graziani2018}
{Graziani} L.,  {Ciardi} B.,   {Glatzle} M.,  2018, \mn@doi [\mnras] {10.1093/mnras/sty1367}, \href {https://ui.adsabs.harvard.edu/abs/2018MNRAS.479.4320G} {479, 4320}

\bibitem[\protect\citeauthoryear{{Greene} et~al.,}{{Greene} et~al.}{2023}]{Greene2023}
{Greene} J.~E.,  et~al., 2023, \mn@doi [arXiv e-prints] {10.48550/arXiv.2309.05714}, \href {https://ui.adsabs.harvard.edu/abs/2023arXiv230905714G} {p. arXiv:2309.05714}

\bibitem[\protect\citeauthoryear{{Greig}, {Mesinger}  \& {Ba{\~n}ados}}{{Greig} et~al.}{2019}]{greig2019}
{Greig} B.,  {Mesinger} A.,   {Ba{\~n}ados} E.,  2019, \mn@doi [\mnras] {10.1093/mnras/stz230}, \href {https://ui.adsabs.harvard.edu/abs/2019MNRAS.484.5094G} {484, 5094}

\bibitem[\protect\citeauthoryear{{Greig} et~al.,}{{Greig} et~al.}{2024}]{Greig2024}
{Greig} B.,  et~al., 2024, \mn@doi [\mnras] {10.1093/mnras/stae1080}, \href {https://ui.adsabs.harvard.edu/abs/2024MNRAS.530.3208G} {530, 3208}

\bibitem[\protect\citeauthoryear{{Grissom}, {Ballantyne}  \& {Wise}}{{Grissom} et~al.}{2014}]{Grissom2014}
{Grissom} R.~L.,  {Ballantyne} D.~R.,   {Wise} J.~H.,  2014, \mn@doi [\aap] {10.1051/0004-6361/201322637}, \href {https://ui.adsabs.harvard.edu/abs/2014A&A...561A..90G} {561, A90}

\bibitem[\protect\citeauthoryear{{Gunn} \& {Peterson}}{{Gunn} \& {Peterson}}{1965}]{gunn1965}
{Gunn} J.~E.,  {Peterson} B.~A.,  1965, \mn@doi [\apj] {10.1086/148444}, \href {https://ui.adsabs.harvard.edu/abs/1965ApJ...142.1633G} {142, 1633}

\bibitem[\protect\citeauthoryear{{Haardt} \& {Madau}}{{Haardt} \& {Madau}}{2012}]{Haardt2012}
{Haardt} F.,  {Madau} P.,  2012, \mn@doi [\apj] {10.1088/0004-637X/746/2/125}, \href {https://ui.adsabs.harvard.edu/abs/2012ApJ...746..125H} {746, 125}

\bibitem[\protect\citeauthoryear{{Hariharan}, {Graziani}, {Ciardi}, {Miniati}  \& {Bungartz}}{{Hariharan} et~al.}{2017}]{hariharan2017}
{Hariharan} N.,  {Graziani} L.,  {Ciardi} B.,  {Miniati} F.,   {Bungartz} H.~J.,  2017, \mn@doi [\mnras] {10.1093/mnras/stx162}, \href {https://ui.adsabs.harvard.edu/abs/2017MNRAS.467.2458H} {467, 2458}

\bibitem[\protect\citeauthoryear{{Harikane} et~al.,}{{Harikane} et~al.}{2023}]{Harikane2023}
{Harikane} Y.,  et~al., 2023, \mn@doi [\apj] {10.3847/1538-4357/ad029e}, \href {https://ui.adsabs.harvard.edu/abs/2023ApJ...959...39H} {959, 39}

\bibitem[\protect\citeauthoryear{{Hassan}, {Dav{\'e}}, {Mitra}, {Finlator}, {Ciardi}  \& {Santos}}{{Hassan} et~al.}{2018}]{hassan2018}
{Hassan} S.,  {Dav{\'e}} R.,  {Mitra} S.,  {Finlator} K.,  {Ciardi} B.,   {Santos} M.~G.,  2018, \mn@doi [\mnras] {10.1093/mnras/stx2194}, \href {https://ui.adsabs.harvard.edu/abs/2018MNRAS.473..227H} {473, 227}

\bibitem[\protect\citeauthoryear{{Hellsten}, {Hernquist}, {Katz}  \& {Weinberg}}{{Hellsten} et~al.}{1998}]{hellsten1998}
{Hellsten} U.,  {Hernquist} L.,  {Katz} N.,   {Weinberg} D.~H.,  1998, \mn@doi [\apj] {10.1086/305622}, \href {https://ui.adsabs.harvard.edu/abs/1998ApJ...499..172H} {499, 172}

\bibitem[\protect\citeauthoryear{{Hoag} et~al.,}{{Hoag} et~al.}{2019}]{hoag2019}
{Hoag} A.,  et~al., 2019, \mn@doi [\apj] {10.3847/1538-4357/ab1de7}, \href {https://ui.adsabs.harvard.edu/abs/2019ApJ...878...12H} {878, 12}

\bibitem[\protect\citeauthoryear{{Hockney} \& {Eastwood}}{{Hockney} \& {Eastwood}}{1988}]{hockney1988}
{Hockney} R.~W.,  {Eastwood} J.~W.,  1988, {Computer simulation using particles}

\bibitem[\protect\citeauthoryear{{Hunter}}{{Hunter}}{2007}]{Hunter2007}
{Hunter} J.~D.,  2007, \mn@doi [Computing in Science and Engineering] {10.1109/MCSE.2007.55}, \href {https://ui.adsabs.harvard.edu/abs/2007CSE.....9...90H} {9, 90}

\bibitem[\protect\citeauthoryear{{Hutter}, {Dayal}, {Yepes}, {Gottl{\"o}ber}, {Legrand}  \& {Ucci}}{{Hutter} et~al.}{2021}]{hutter2021}
{Hutter} A.,  {Dayal} P.,  {Yepes} G.,  {Gottl{\"o}ber} S.,  {Legrand} L.,   {Ucci} G.,  2021, \mn@doi [\mnras] {10.1093/mnras/stab602}, \href {https://ui.adsabs.harvard.edu/abs/2021MNRAS.503.3698H} {503, 3698}

\bibitem[\protect\citeauthoryear{{Jin} et~al.,}{{Jin} et~al.}{2023}]{Jin2023}
{Jin} X.,  et~al., 2023, \mn@doi [\apj] {10.3847/1538-4357/aca678}, \href {https://ui.adsabs.harvard.edu/abs/2023ApJ...942...59J} {942, 59}

\bibitem[\protect\citeauthoryear{Jones, Oliphant  \& Peterson}{Jones et~al.}{2001}]{Jones2001}
Jones E.,  Oliphant T.,   Peterson P.,  2001

\bibitem[\protect\citeauthoryear{{Jung} et~al.,}{{Jung} et~al.}{2022}]{jung2022}
{Jung} I.,  et~al., 2022, \mn@doi [arXiv e-prints] {10.48550/arXiv.2212.09850}, \href {https://ui.adsabs.harvard.edu/abs/2022arXiv221209850J} {p. arXiv:2212.09850}

\bibitem[\protect\citeauthoryear{{Juod{\v{z}}balis} et~al.,}{{Juod{\v{z}}balis} et~al.}{2023}]{Juodzbalis2023}
{Juod{\v{z}}balis} I.,  et~al., 2023, \mn@doi [\mnras] {10.1093/mnras/stad2396}, \href {https://ui.adsabs.harvard.edu/abs/2023MNRAS.525.1353J} {525, 1353}

\bibitem[\protect\citeauthoryear{{Kakiichi} et~al.,}{{Kakiichi} et~al.}{2018}]{kakiichi2018}
{Kakiichi} K.,  et~al., 2018, \mn@doi [\mnras] {10.1093/mnras/sty1318}, \href {https://ui.adsabs.harvard.edu/abs/2018MNRAS.479...43K} {479, 43}

\bibitem[\protect\citeauthoryear{{Kakiichi} et~al.,}{{Kakiichi} et~al.}{2025}]{koki2025}
{Kakiichi} K.,  et~al., 2025, \mn@doi [arXiv e-prints] {10.48550/arXiv.2503.07074}, \href {https://ui.adsabs.harvard.edu/abs/2025arXiv250307074K} {p. arXiv:2503.07074}

\bibitem[\protect\citeauthoryear{{Kamran}, {Ghara}, {Majumdar}, {Mellema}, {Bharadwaj}, {Pritchard}, {Mondal}  \& {Iliev}}{{Kamran} et~al.}{2022}]{kamran2022}
{Kamran} M.,  {Ghara} R.,  {Majumdar} S.,  {Mellema} G.,  {Bharadwaj} S.,  {Pritchard} J.~R.,  {Mondal} R.,   {Iliev} I.~T.,  2022, \mn@doi [\jcap] {10.1088/1475-7516/2022/11/001}, \href {https://ui.adsabs.harvard.edu/abs/2022JCAP...11..001K} {2022, 001}

\bibitem[\protect\citeauthoryear{{Kannan}, {Garaldi}, {Smith}, {Pakmor}, {Springel}, {Vogelsberger}  \& {Hernquist}}{{Kannan} et~al.}{2022}]{kannan2022}
{Kannan} R.,  {Garaldi} E.,  {Smith} A.,  {Pakmor} R.,  {Springel} V.,  {Vogelsberger} M.,   {Hernquist} L.,  2022, \mn@doi [\mnras] {10.1093/mnras/stab3710}, \href {https://ui.adsabs.harvard.edu/abs/2022MNRAS.511.4005K} {511, 4005}

\bibitem[\protect\citeauthoryear{{Kara{\c{c}}ayl{\i}}, {Font-Ribera}  \& {Padmanabhan}}{{Kara{\c{c}}ayl{\i}} et~al.}{2020}]{Karacayli2020}
{Kara{\c{c}}ayl{\i}} N.~G.,  {Font-Ribera} A.,   {Padmanabhan} N.,  2020, \mn@doi [\mnras] {10.1093/mnras/staa2331}, \href {https://ui.adsabs.harvard.edu/abs/2020MNRAS.497.4742K} {497, 4742}

\bibitem[\protect\citeauthoryear{{Kashino}, {Lilly}, {Matthee}, {Eilers}, {Mackenzie}, {Bordoloi}  \& {Simcoe}}{{Kashino} et~al.}{2023}]{kashino2023}
{Kashino} D.,  {Lilly} S.~J.,  {Matthee} J.,  {Eilers} A.-C.,  {Mackenzie} R.,  {Bordoloi} R.,   {Simcoe} R.~A.,  2023, \mn@doi [\apj] {10.3847/1538-4357/acc588}, \href {https://ui.adsabs.harvard.edu/abs/2023ApJ...950...66K} {950, 66}

\bibitem[\protect\citeauthoryear{{Kashino} et~al.,}{{Kashino} et~al.}{2025}]{Kashino2025}
{Kashino} D.,  et~al., 2025, \mn@doi [arXiv e-prints] {10.48550/arXiv.2506.03121}, \href {https://ui.adsabs.harvard.edu/abs/2025arXiv250603121K} {p. arXiv:2506.03121}

\bibitem[\protect\citeauthoryear{{Keating}, {Puchwein}, {Haehnelt}, {Bird}  \& {Bolton}}{{Keating} et~al.}{2016}]{Keating2016}
{Keating} L.~C.,  {Puchwein} E.,  {Haehnelt} M.~G.,  {Bird} S.,   {Bolton} J.~S.,  2016, \mn@doi [\mnras] {10.1093/mnras/stw1306}, \href {https://ui.adsabs.harvard.edu/abs/2016MNRAS.461..606K} {461, 606}

\bibitem[\protect\citeauthoryear{{Komatsu} et~al.,}{{Komatsu} et~al.}{2011}]{Komatsu2011}
{Komatsu} E.,  et~al., 2011, \mn@doi [\apjs] {10.1088/0067-0049/192/2/18}, \href {https://ui.adsabs.harvard.edu/abs/2011ApJS..192...18K} {192, 18}

\bibitem[\protect\citeauthoryear{{Kulkarni}, {Keating}, {Haehnelt}, {Bosman}, {Puchwein}, {Chardin}  \& {Aubert}}{{Kulkarni} et~al.}{2019}]{kulkarni2019}
{Kulkarni} G.,  {Keating} L.~C.,  {Haehnelt} M.~G.,  {Bosman} S. E.~I.,  {Puchwein} E.,  {Chardin} J.,   {Aubert} D.,  2019, \mn@doi [\mnras] {10.1093/mnrasl/slz025}, \href {https://ui.adsabs.harvard.edu/abs/2019MNRAS.485L..24K} {485, L24}

\bibitem[\protect\citeauthoryear{{Lamberts}, {Puchwein}, {Pfrommer}, {Chang}, {Shalaby}, {Broderick}, {Tiede}  \& {Rudie}}{{Lamberts} et~al.}{2022}]{lamberts2022}
{Lamberts} A.,  {Puchwein} E.,  {Pfrommer} C.,  {Chang} P.,  {Shalaby} M.,  {Broderick} A.,  {Tiede} P.,   {Rudie} G.,  2022, \mn@doi [\mnras] {10.1093/mnras/stac553}, \href {https://ui.adsabs.harvard.edu/abs/2022MNRAS.512.3045L} {512, 3045}

\bibitem[\protect\citeauthoryear{{Larson} et~al.,}{{Larson} et~al.}{2023}]{Larson2023}
{Larson} R.~L.,  et~al., 2023, \mn@doi [\apjl] {10.3847/2041-8213/ace619}, \href {https://ui.adsabs.harvard.edu/abs/2023ApJ...953L..29L} {953, L29}

\bibitem[\protect\citeauthoryear{{Leite}, {Evoli}, {D'Angelo}, {Ciardi}, {Sigl}  \& {Ferrara}}{{Leite} et~al.}{2017}]{leite2017}
{Leite} N.,  {Evoli} C.,  {D'Angelo} M.,  {Ciardi} B.,  {Sigl} G.,   {Ferrara} A.,  2017, \mn@doi [\mnras] {10.1093/mnras/stx805}, \href {https://ui.adsabs.harvard.edu/abs/2017MNRAS.469..416L} {469, 416}

\bibitem[\protect\citeauthoryear{{Liu}, {Slatyer}  \& {Zavala}}{{Liu} et~al.}{2016}]{liu2016}
{Liu} H.,  {Slatyer} T.~R.,   {Zavala} J.,  2016, \mn@doi [\prd] {10.1103/PhysRevD.94.063507}, \href {https://ui.adsabs.harvard.edu/abs/2016PhRvD..94f3507L} {94, 063507}

\bibitem[\protect\citeauthoryear{Loeb \& Barkana}{Loeb \& Barkana}{2001}]{loeb2001}
Loeb A.,  Barkana R.,  2001, \mn@doi [Annual Review of Astronomy and Astrophysics] {10.1146/annurev.astro.39.1.19}, 39, 19

\bibitem[\protect\citeauthoryear{{Ma}, {Ciardi}, {Kakiichi}, {Zaroubi}, {Zhi}  \& {Busch}}{{Ma} et~al.}{2020}]{qingbo2020}
{Ma} Q.-B.,  {Ciardi} B.,  {Kakiichi} K.,  {Zaroubi} S.,  {Zhi} Q.-J.,   {Busch} P.,  2020, \mn@doi [\apj] {10.3847/1538-4357/ab5b95}, \href {https://ui.adsabs.harvard.edu/abs/2020ApJ...888..112M} {888, 112}

\bibitem[\protect\citeauthoryear{{Ma}, {Ciardi}, {Eide}, {Busch}, {Mao}  \& {Zhi}}{{Ma} et~al.}{2021}]{qingbo2021}
{Ma} Q.-B.,  {Ciardi} B.,  {Eide} M.~B.,  {Busch} P.,  {Mao} Y.,   {Zhi} Q.-J.,  2021, \mn@doi [\apj] {10.3847/1538-4357/abefd5}, \href {https://ui.adsabs.harvard.edu/abs/2021ApJ...912..143M} {912, 143}

\bibitem[\protect\citeauthoryear{{Ma}, {Fiaschi}, {Ciardi}, {Busch}  \& {Eide}}{{Ma} et~al.}{2022}]{qingbo2022}
{Ma} Q.-B.,  {Fiaschi} S.,  {Ciardi} B.,  {Busch} P.,   {Eide} M.~B.,  2022, \mn@doi [\mnras] {10.1093/mnras/stac1018}, \href {https://ui.adsabs.harvard.edu/abs/2022MNRAS.513.1513M} {513, 1513}

\bibitem[\protect\citeauthoryear{Madau}{Madau}{2017}]{Madau2017}
Madau P.,  2017, \mn@doi [The Astrophysical Journal] {10.3847/1538-4357/aa9715}, 851, 50

\bibitem[\protect\citeauthoryear{{Madau} \& {Fragos}}{{Madau} \& {Fragos}}{2017}]{madaufragos2017}
{Madau} P.,  {Fragos} T.,  2017, \mn@doi [\apj] {10.3847/1538-4357/aa6af9}, \href {https://ui.adsabs.harvard.edu/abs/2017ApJ...840...39M} {840, 39}

\bibitem[\protect\citeauthoryear{{Madau}, {Giallongo}, {Grazian}  \& {Haardt}}{{Madau} et~al.}{2024}]{madau2024}
{Madau} P.,  {Giallongo} E.,  {Grazian} A.,   {Haardt} F.,  2024, \mn@doi [\apj] {10.3847/1538-4357/ad5ce8}, \href {https://ui.adsabs.harvard.edu/abs/2024ApJ...971...75M} {971, 75}

\bibitem[\protect\citeauthoryear{{Maiolino} et~al.,}{{Maiolino} et~al.}{2023}]{Maiolino2023}
{Maiolino} R.,  et~al., 2023, \mn@doi [arXiv e-prints] {10.48550/arXiv.2308.01230}, \href {https://ui.adsabs.harvard.edu/abs/2023arXiv230801230M} {p. arXiv:2308.01230}

\bibitem[\protect\citeauthoryear{{Maselli}, {Ferrara}  \& {Ciardi}}{{Maselli} et~al.}{2003}]{maselli2003}
{Maselli} A.,  {Ferrara} A.,   {Ciardi} B.,  2003, \mn@doi [\mnras] {10.1046/j.1365-8711.2003.06979.x}, \href {https://ui.adsabs.harvard.edu/abs/2003MNRAS.345..379M} {345, 379}

\bibitem[\protect\citeauthoryear{{Maselli}, {Ciardi}  \& {Kanekar}}{{Maselli} et~al.}{2009}]{maselli2009}
{Maselli} A.,  {Ciardi} B.,   {Kanekar} A.,  2009, \mn@doi [\mnras] {10.1111/j.1365-2966.2008.14197.x}, \href {https://ui.adsabs.harvard.edu/abs/2009MNRAS.393..171M} {393, 171}

\bibitem[\protect\citeauthoryear{Mason, Treu, Dijkstra, Mesinger, Trenti, Pentericci, de Barros  \& Vanzella}{Mason et~al.}{2018}]{Mason2018}
Mason C.~A.,  Treu T.,  Dijkstra M.,  Mesinger A.,  Trenti M.,  Pentericci L.,  de Barros S.,   Vanzella E.,  2018, \mn@doi [The Astrophysical Journal] {10.3847/1538-4357/aab0a7}, 856, 2

\bibitem[\protect\citeauthoryear{{Mason}, {Chen}, {Stark}, {Lu}, {Topping}  \& {Tang}}{{Mason} et~al.}{2025}]{Mason2025}
{Mason} C.~A.,  {Chen} Z.,  {Stark} D.~P.,  {Lu} T.-Y.,  {Topping} M.,   {Tang} M.,  2025, \mn@doi [arXiv e-prints] {10.48550/arXiv.2501.11702}, \href {https://ui.adsabs.harvard.edu/abs/2025arXiv250111702M} {p. arXiv:2501.11702}

\bibitem[\protect\citeauthoryear{{Matthee}, {Sobral}, {Gronke}, {Paulino-Afonso}, {Stefanon}  \& {R{\"o}ttgering}}{{Matthee} et~al.}{2018}]{matthee2018}
{Matthee} J.,  {Sobral} D.,  {Gronke} M.,  {Paulino-Afonso} A.,  {Stefanon} M.,   {R{\"o}ttgering} H.,  2018, \mn@doi [\aap] {10.1051/0004-6361/201833528}, \href {https://ui.adsabs.harvard.edu/abs/2018A&A...619A.136M} {619, A136}

\bibitem[\protect\citeauthoryear{Matthee et~al.,}{Matthee et~al.}{2022}]{Matthee2022}
Matthee J.,  et~al., 2022, \mn@doi [Monthly Notices of the Royal Astronomical Society] {10.1093/mnras/stac801}

\bibitem[\protect\citeauthoryear{{McGreer}, {Mesinger}  \& {D'Odorico}}{{McGreer} et~al.}{2015}]{McGreer2015}
{McGreer} I.~D.,  {Mesinger} A.,   {D'Odorico} V.,  2015, \mn@doi [\mnras] {10.1093/mnras/stu2449}, \href {https://ui.adsabs.harvard.edu/abs/2015MNRAS.447..499M} {447, 499}

\bibitem[\protect\citeauthoryear{{McGreer}, {Fan}, {Jiang}  \& {Cai}}{{McGreer} et~al.}{2018}]{McGreer2018}
{McGreer} I.~D.,  {Fan} X.,  {Jiang} L.,   {Cai} Z.,  2018, \mn@doi [\aj] {10.3847/1538-3881/aaaab4}, \href {https://ui.adsabs.harvard.edu/abs/2018AJ....155..131M} {155, 131}

\bibitem[\protect\citeauthoryear{{Mesinger} \& {Furlanetto}}{{Mesinger} \& {Furlanetto}}{2007}]{mesinger2007}
{Mesinger} A.,  {Furlanetto} S.,  2007, \mn@doi [\apj] {10.1086/521806}, \href {https://ui.adsabs.harvard.edu/abs/2007ApJ...669..663M} {669, 663}

\bibitem[\protect\citeauthoryear{{Meyer}, {Bosman}, {Kakiichi}  \& {Ellis}}{{Meyer} et~al.}{2019}]{meyer2019}
{Meyer} R.~A.,  {Bosman} S. E.~I.,  {Kakiichi} K.,   {Ellis} R.~S.,  2019, \mn@doi [\mnras] {10.1093/mnras/sty2954}, \href {https://ui.adsabs.harvard.edu/abs/2019MNRAS.483...19M} {483, 19}

\bibitem[\protect\citeauthoryear{{Meyer} et~al.,}{{Meyer} et~al.}{2020}]{meyer2020}
{Meyer} R.~A.,  et~al., 2020, \mn@doi [\mnras] {10.1093/mnras/staa746}, \href {https://ui.adsabs.harvard.edu/abs/2020MNRAS.494.1560M} {494, 1560}

\bibitem[\protect\citeauthoryear{{Mirabel}, {Dijkstra}, {Laurent}, {Loeb}  \& {Pritchard}}{{Mirabel} et~al.}{2011}]{mirabel2011}
{Mirabel} I.~F.,  {Dijkstra} M.,  {Laurent} P.,  {Loeb} A.,   {Pritchard} J.~R.,  2011, \mn@doi [\aap] {10.1051/0004-6361/201016357}, \href {https://ui.adsabs.harvard.edu/abs/2011A&A...528A.149M} {528, A149}

\bibitem[\protect\citeauthoryear{{Molaro} et~al.,}{{Molaro} et~al.}{2022}]{Molaro2022}
{Molaro} M.,  et~al., 2022, \mn@doi [\mnras] {10.1093/mnras/stab3416}, \href {https://ui.adsabs.harvard.edu/abs/2022MNRAS.509.6119M} {509, 6119}

\bibitem[\protect\citeauthoryear{{Monaghan}}{{Monaghan}}{1992}]{monaghan1992}
{Monaghan} J.~J.,  1992, \mn@doi [\araa] {10.1146/annurev.aa.30.090192.002551}, \href {https://ui.adsabs.harvard.edu/abs/1992ARA&A..30..543M} {30, 543}

\bibitem[\protect\citeauthoryear{{Naidu}, {Forrest}, {Oesch}, {Tran}  \& {Holden}}{{Naidu} et~al.}{2018}]{naidu2018}
{Naidu} R.~P.,  {Forrest} B.,  {Oesch} P.~A.,  {Tran} K.-V.~H.,   {Holden} B.~P.,  2018, \mn@doi [\mnras] {10.1093/mnras/sty961}, \href {https://ui.adsabs.harvard.edu/abs/2018MNRAS.478..791N} {478, 791}

\bibitem[\protect\citeauthoryear{Naidu, Tacchella, Mason, Bose, Oesch  \& Conroy}{Naidu et~al.}{2020}]{Naidu2020}
Naidu R.~P.,  Tacchella S.,  Mason C.~A.,  Bose S.,  Oesch P.~A.,   Conroy C.,  2020, \mn@doi [The Astrophysical Journal] {10.3847/1538-4357/ab7cc9}, 892, 109

\bibitem[\protect\citeauthoryear{Naidu et~al.,}{Naidu et~al.}{2021}]{Naidu2021}
Naidu R.~P.,  et~al., 2021, \mn@doi [Monthly Notices of the Royal Astronomical Society] {10.1093/mnras/stab3601}, 510, 4582

\bibitem[\protect\citeauthoryear{{Nakane} et~al.,}{{Nakane} et~al.}{2024}]{nakane2024}
{Nakane} M.,  et~al., 2024, \mn@doi [\apj] {10.3847/1538-4357/ad38c2}, \href {https://ui.adsabs.harvard.edu/abs/2024ApJ...967...28N} {967, 28}

\bibitem[\protect\citeauthoryear{{Napolitano} et~al.,}{{Napolitano} et~al.}{2024}]{napolitano2024}
{Napolitano} L.,  et~al., 2024, \mn@doi [\aap] {10.1051/0004-6361/202449644}, \href {https://ui.adsabs.harvard.edu/abs/2024A&A...688A.106N} {688, A106}

\bibitem[\protect\citeauthoryear{{Nasir} \& {D'Aloisio}}{{Nasir} \& {D'Aloisio}}{2020}]{nasir2020}
{Nasir} F.,  {D'Aloisio} A.,  2020, \mn@doi [\mnras] {10.1093/mnras/staa894}, \href {https://ui.adsabs.harvard.edu/abs/2020MNRAS.494.3080N} {494, 3080}

\bibitem[\protect\citeauthoryear{{Nasir}, {Bolton}, {Viel}, {Kim}, {Haehnelt}, {Puchwein}  \& {Sijacki}}{{Nasir} et~al.}{2017}]{Nasir2017}
{Nasir} F.,  {Bolton} J.~S.,  {Viel} M.,  {Kim} T.-S.,  {Haehnelt} M.~G.,  {Puchwein} E.,   {Sijacki} D.,  2017, \mn@doi [\mnras] {10.1093/mnras/stx1648}, \href {https://ui.adsabs.harvard.edu/abs/2017MNRAS.471.1056N} {471, 1056}

\bibitem[\protect\citeauthoryear{{Nath} \& {Biermann}}{{Nath} \& {Biermann}}{1993}]{nath1993}
{Nath} B.~B.,  {Biermann} P.~L.,  1993, \mn@doi [\mnras] {10.1093/mnras/265.1.241}, \href {https://ui.adsabs.harvard.edu/abs/1993MNRAS.265..241N} {265, 241}

\bibitem[\protect\citeauthoryear{{Ning} et~al.,}{{Ning} et~al.}{2024}]{ning2024}
{Ning} Y.,  et~al., 2024, \mn@doi [\apjl] {10.3847/2041-8213/ad292f}, \href {https://ui.adsabs.harvard.edu/abs/2024ApJ...963L..38N} {963, L38}

\bibitem[\protect\citeauthoryear{{Noble}, {Kamran}, {Majumdar}, {Murmu}, {Ghara}, {Mellema}, {Iliev}  \& {Pritchard}}{{Noble} et~al.}{2024}]{noble2024}
{Noble} L.,  {Kamran} M.,  {Majumdar} S.,  {Murmu} C.~S.,  {Ghara} R.,  {Mellema} G.,  {Iliev} I.~T.,   {Pritchard} J.~R.,  2024, \mn@doi [\jcap] {10.1088/1475-7516/2024/10/003}, \href {https://ui.adsabs.harvard.edu/abs/2024JCAP...10..003N} {2024, 003}

\bibitem[\protect\citeauthoryear{{Ocvirk} et~al.,}{{Ocvirk} et~al.}{2020}]{ockvirk2020}
{Ocvirk} P.,  et~al., 2020, \mn@doi [\mnras] {10.1093/mnras/staa1266}, \href {https://ui.adsabs.harvard.edu/abs/2020MNRAS.496.4087O} {496, 4087}

\bibitem[\protect\citeauthoryear{{Owen}, {Wu}, {Jin}, {Surajbali}  \& {Kataoka}}{{Owen} et~al.}{2019}]{owen2019}
{Owen} E.~R.,  {Wu} K.,  {Jin} X.,  {Surajbali} P.,   {Kataoka} N.,  2019, \mn@doi [\aap] {10.1051/0004-6361/201834350}, \href {https://ui.adsabs.harvard.edu/abs/2019A&A...626A..85O} {626, A85}

\bibitem[\protect\citeauthoryear{{Pagano}, {Delouis}, {Mottet}, {Puget}  \& {Vibert}}{{Pagano} et~al.}{2020}]{pagano2020}
{Pagano} L.,  {Delouis} J.~M.,  {Mottet} S.,  {Puget} J.~L.,   {Vibert} L.,  2020, \mn@doi [\aap] {10.1051/0004-6361/201936630}, \href {https://ui.adsabs.harvard.edu/abs/2020A&A...635A..99P} {635, A99}

\bibitem[\protect\citeauthoryear{{Parsa}, {Dunlop}  \& {McLure}}{{Parsa} et~al.}{2018}]{Parsha2018}
{Parsa} S.,  {Dunlop} J.~S.,   {McLure} R.~J.,  2018, \mn@doi [\mnras] {10.1093/mnras/stx2887}, \href {https://ui.adsabs.harvard.edu/abs/2018MNRAS.474.2904P} {474, 2904}

\bibitem[\protect\citeauthoryear{Pentericci et~al.,}{Pentericci et~al.}{2014}]{Pentericci2014}
Pentericci L.,  et~al., 2014, \mn@doi [The Astrophysical Journal] {10.1088/0004-637x/793/2/113}, 793, 113

\bibitem[\protect\citeauthoryear{{Planck Collaboration} et~al.,}{{Planck Collaboration} et~al.}{2014}]{planck2013}
{Planck Collaboration} et~al., 2014, \mn@doi [A\&A] {10.1051/0004-6361/201321591}, 571, A16

\bibitem[\protect\citeauthoryear{{Planck Collaboration} et~al.,}{{Planck Collaboration} et~al.}{2016}]{planck2016}
{Planck Collaboration} et~al., 2016, \mn@doi [\aap] {10.1051/0004-6361/201628897}, \href {https://ui.adsabs.harvard.edu/abs/2016A&A...596A.108P} {596, A108}

\bibitem[\protect\citeauthoryear{{Puchwein}, {Pfrommer}, {Springel}, {Broderick}  \& {Chang}}{{Puchwein} et~al.}{2012}]{puchwein2012}
{Puchwein} E.,  {Pfrommer} C.,  {Springel} V.,  {Broderick} A.~E.,   {Chang} P.,  2012, \mn@doi [\mnras] {10.1111/j.1365-2966.2012.20738.x}, \href {https://ui.adsabs.harvard.edu/abs/2012MNRAS.423..149P} {423, 149}

\bibitem[\protect\citeauthoryear{Raskutti, Bolton, Wyithe  \& Becker}{Raskutti et~al.}{2012}]{Raskutti2012}
Raskutti S.,  Bolton J.~S.,  Wyithe J. S.~B.,   Becker G.~D.,  2012, \mn@doi [Monthly Notices of the Royal Astronomical Society] {10.1111/j.1365-2966.2011.20401.x}, 421, 1969

\bibitem[\protect\citeauthoryear{{Robertson}, {Ellis}, {Furlanetto}  \& {Dunlop}}{{Robertson} et~al.}{2015}]{Robertson2015}
{Robertson} B.~E.,  {Ellis} R.~S.,  {Furlanetto} S.~R.,   {Dunlop} J.~S.,  2015, \mn@doi [\apjl] {10.1088/2041-8205/802/2/L19}, \href {https://ui.adsabs.harvard.edu/abs/2015ApJ...802L..19R} {802, L19}

\bibitem[\protect\citeauthoryear{{Rosdahl} et~al.,}{{Rosdahl} et~al.}{2018}]{rosdahl2018}
{Rosdahl} J.,  et~al., 2018, \mn@doi [\mnras] {10.1093/mnras/sty1655}, \href {https://ui.adsabs.harvard.edu/abs/2018MNRAS.479..994R} {479, 994}

\bibitem[\protect\citeauthoryear{{Ross}, {Dixon}, {Iliev}  \& {Mellema}}{{Ross} et~al.}{2017}]{ross2017}
{Ross} H.~E.,  {Dixon} K.~L.,  {Iliev} I.~T.,   {Mellema} G.,  2017, \mn@doi [\mnras] {10.1093/mnras/stx649}, \href {https://ui.adsabs.harvard.edu/abs/2017MNRAS.468.3785R} {468, 3785}

\bibitem[\protect\citeauthoryear{{Santos}, {Ferramacho}, {Silva}, {Amblard}  \& {Cooray}}{{Santos} et~al.}{2010}]{santos2010}
{Santos} M.~G.,  {Ferramacho} L.,  {Silva} M.~B.,  {Amblard} A.,   {Cooray} A.,  2010, \mn@doi [\mnras] {10.1111/j.1365-2966.2010.16898.x}, \href {https://ui.adsabs.harvard.edu/abs/2010MNRAS.406.2421S} {406, 2421}

\bibitem[\protect\citeauthoryear{{Saxena} et~al.,}{{Saxena} et~al.}{2024}]{saxena2024}
{Saxena} A.,  et~al., 2024, \mn@doi [\aap] {10.1051/0004-6361/202347132}, \href {https://ui.adsabs.harvard.edu/abs/2024A&A...684A..84S} {684, A84}

\bibitem[\protect\citeauthoryear{{Sazonov} \& {Khabibullin}}{{Sazonov} \& {Khabibullin}}{2017}]{sazonov2017}
{Sazonov} S.~Y.,  {Khabibullin} I.~I.,  2017, \mn@doi [Astronomy Letters] {10.1134/S1063773717040077}, \href {https://ui.adsabs.harvard.edu/abs/2017AstL...43..211S} {43, 211}

\bibitem[\protect\citeauthoryear{{Sazonov} \& {Sunyaev}}{{Sazonov} \& {Sunyaev}}{2015}]{sazonov2015}
{Sazonov} S.,  {Sunyaev} R.,  2015, \mn@doi [\mnras] {10.1093/mnras/stv2255}, \href {https://ui.adsabs.harvard.edu/abs/2015MNRAS.454.3464S} {454, 3464}

\bibitem[\protect\citeauthoryear{{Spina}, {Bosman}, {Davies}, {Gaikwad}  \& {Zhu}}{{Spina} et~al.}{2024}]{spina2024}
{Spina} B.,  {Bosman} S. E.~I.,  {Davies} F.~B.,  {Gaikwad} P.,   {Zhu} Y.,  2024, \mn@doi [\aap] {10.1051/0004-6361/202450798}, \href {https://ui.adsabs.harvard.edu/abs/2024A&A...688L..26S} {688, L26}

\bibitem[\protect\citeauthoryear{{Springel}}{{Springel}}{2005}]{gadget2}
{Springel} V.,  2005, \mn@doi [\mnras] {10.1111/j.1365-2966.2005.09655.x}, \href {https://ui.adsabs.harvard.edu/abs/2005MNRAS.364.1105S} {364, 1105}

\bibitem[\protect\citeauthoryear{{Springel}, {Yoshida}  \& {White}}{{Springel} et~al.}{2001a}]{Volker2001}
{Springel} V.,  {Yoshida} N.,   {White} S. D.~M.,  2001a, \mn@doi [\na] {10.1016/S1384-1076(01)00042-2}, \href {https://ui.adsabs.harvard.edu/abs/2001NewA....6...79S} {6, 79}

\bibitem[\protect\citeauthoryear{{Springel}, {White}, {Tormen}  \& {Kauffmann}}{{Springel} et~al.}{2001b}]{springel2001}
{Springel} V.,  {White} S. D.~M.,  {Tormen} G.,   {Kauffmann} G.,  2001b, \mn@doi [\mnras] {10.1046/j.1365-8711.2001.04912.x}, \href {https://ui.adsabs.harvard.edu/abs/2001MNRAS.328..726S} {328, 726}

\bibitem[\protect\citeauthoryear{{Steidel}, {Bogosavljevi{\'c}}, {Shapley}, {Reddy}, {Rudie}, {Pettini}, {Trainor}  \& {Strom}}{{Steidel} et~al.}{2018}]{steidel2018}
{Steidel} C.~C.,  {Bogosavljevi{\'c}} M.,  {Shapley} A.~E.,  {Reddy} N.~A.,  {Rudie} G.~C.,  {Pettini} M.,  {Trainor} R.~F.,   {Strom} A.~L.,  2018, \mn@doi [\apj] {10.3847/1538-4357/aaed28}, \href {https://ui.adsabs.harvard.edu/abs/2018ApJ...869..123S} {869, 123}

\bibitem[\protect\citeauthoryear{{Strickland} \& {Stevens}}{{Strickland} \& {Stevens}}{2000}]{strickland2000}
{Strickland} D.~K.,  {Stevens} I.~R.,  2000, \mn@doi [\mnras] {10.1046/j.1365-8711.2000.03391.x}, \href {https://ui.adsabs.harvard.edu/abs/2000MNRAS.314..511S} {314, 511}

\bibitem[\protect\citeauthoryear{{Suchkov}, {Balsara}, {Heckman}  \& {Leitherer}}{{Suchkov} et~al.}{1994}]{suchkov1994}
{Suchkov} A.~A.,  {Balsara} D.~S.,  {Heckman} T.~M.,   {Leitherer} C.,  1994, \mn@doi [\apj] {10.1086/174427}, \href {https://ui.adsabs.harvard.edu/abs/1994ApJ...430..511S} {430, 511}

\bibitem[\protect\citeauthoryear{{Tang}, {Stark}, {Topping}, {Mason}  \& {Ellis}}{{Tang} et~al.}{2024}]{tang2024}
{Tang} M.,  {Stark} D.~P.,  {Topping} M.~W.,  {Mason} C.,   {Ellis} R.~S.,  2024, \mn@doi [arXiv e-prints] {10.48550/arXiv.2408.01507}, \href {https://ui.adsabs.harvard.edu/abs/2024arXiv240801507T} {p. arXiv:2408.01507}

\bibitem[\protect\citeauthoryear{{Tepper-Garc{\'\i}a}}{{Tepper-Garc{\'\i}a}}{2006}]{tepper2006}
{Tepper-Garc{\'\i}a} T.,  2006, \mn@doi [\mnras] {10.1111/j.1365-2966.2006.10450.x}, \href {https://ui.adsabs.harvard.edu/abs/2006MNRAS.369.2025T} {369, 2025}

\bibitem[\protect\citeauthoryear{{Trebitsch} et~al.,}{{Trebitsch} et~al.}{2021}]{trebitsch2021}
{Trebitsch} M.,  et~al., 2021, \mn@doi [\aap] {10.1051/0004-6361/202037698}, \href {https://ui.adsabs.harvard.edu/abs/2021A&A...653A.154T} {653, A154}

\bibitem[\protect\citeauthoryear{{Umeda}, {Ouchi}, {Nakajima}, {Harikane}, {Ono}, {Xu}, {Isobe}  \& {Zhang}}{{Umeda} et~al.}{2024}]{umeda2024}
{Umeda} H.,  {Ouchi} M.,  {Nakajima} K.,  {Harikane} Y.,  {Ono} Y.,  {Xu} Y.,  {Isobe} Y.,   {Zhang} Y.,  2024, \mn@doi [\apj] {10.3847/1538-4357/ad554e}, \href {https://ui.adsabs.harvard.edu/abs/2024ApJ...971..124U} {971, 124}

\bibitem[\protect\citeauthoryear{{Vanzella} et~al.,}{{Vanzella} et~al.}{2016}]{vanzella2016}
{Vanzella} E.,  et~al., 2016, \mn@doi [\apj] {10.3847/0004-637X/825/1/41}, \href {https://ui.adsabs.harvard.edu/abs/2016ApJ...825...41V} {825, 41}

\bibitem[\protect\citeauthoryear{{Vanzella} et~al.,}{{Vanzella} et~al.}{2018}]{vanzella2018}
{Vanzella} E.,  et~al., 2018, \mn@doi [\mnras] {10.1093/mnrasl/sly023}, \href {https://ui.adsabs.harvard.edu/abs/2018MNRAS.476L..15V} {476, L15}

\bibitem[\protect\citeauthoryear{{Venkatesan} \& {Benson}}{{Venkatesan} \& {Benson}}{2011}]{Venkatesan2011}
{Venkatesan} A.,  {Benson} A.,  2011, \mn@doi [\mnras] {10.1111/j.1365-2966.2011.19407.x}, \href {https://ui.adsabs.harvard.edu/abs/2011MNRAS.417.2264V} {417, 2264}

\bibitem[\protect\citeauthoryear{{Viel}, {Haehnelt}  \& {Springel}}{{Viel} et~al.}{2004}]{Viel2004}
{Viel} M.,  {Haehnelt} M.~G.,   {Springel} V.,  2004, \mn@doi [\mnras] {10.1111/j.1365-2966.2004.08224.x}, \href {https://ui.adsabs.harvard.edu/abs/2004MNRAS.354..684V} {354, 684}

\bibitem[\protect\citeauthoryear{{Viel}, {Schaye}  \& {Booth}}{{Viel} et~al.}{2013}]{Viel2013}
{Viel} M.,  {Schaye} J.,   {Booth} C.~M.,  2013, \mn@doi [\mnras] {10.1093/mnras/sts465}, \href {https://ui.adsabs.harvard.edu/abs/2013MNRAS.429.1734V} {429, 1734}

\bibitem[\protect\citeauthoryear{{Weigel}, {Schawinski}, {Treister}, {Urry}, {Koss}  \& {Trakhtenbrot}}{{Weigel} et~al.}{2015}]{Weigel2015}
{Weigel} A.~K.,  {Schawinski} K.,  {Treister} E.,  {Urry} C.~M.,  {Koss} M.,   {Trakhtenbrot} B.,  2015, \mn@doi [\mnras] {10.1093/mnras/stv184}, \href {https://ui.adsabs.harvard.edu/abs/2015MNRAS.448.3167W} {448, 3167}

\bibitem[\protect\citeauthoryear{{Whitler}, {Mason}, {Ren}, {Dijkstra}, {Mesinger}, {Pentericci}, {Trenti}  \& {Treu}}{{Whitler} et~al.}{2020}]{Whitler2020}
{Whitler} L.~R.,  {Mason} C.~A.,  {Ren} K.,  {Dijkstra} M.,  {Mesinger} A.,  {Pentericci} L.,  {Trenti} M.,   {Treu} T.,  2020, \mn@doi [\mnras] {10.1093/mnras/staa1178}, \href {https://ui.adsabs.harvard.edu/abs/2020MNRAS.495.3602W} {495, 3602}

\bibitem[\protect\citeauthoryear{{Wood} \& {Loeb}}{{Wood} \& {Loeb}}{2000}]{wood2000}
{Wood} K.,  {Loeb} A.,  2000, \mn@doi [\apj] {10.1086/317775}, \href {https://ui.adsabs.harvard.edu/abs/2000ApJ...545...86W} {545, 86}

\bibitem[\protect\citeauthoryear{{Wyithe} \& {Bolton}}{{Wyithe} \& {Bolton}}{2011}]{wyithe2011}
{Wyithe} J. S.~B.,  {Bolton} J.~S.,  2011, \mn@doi [\mnras] {10.1111/j.1365-2966.2010.18030.x}, \href {https://ui.adsabs.harvard.edu/abs/2011MNRAS.412.1926W} {412, 1926}

\bibitem[\protect\citeauthoryear{{Yang} et~al.,}{{Yang} et~al.}{2020}]{yang2020}
{Yang} J.,  et~al., 2020, \mn@doi [\apj] {10.3847/1538-4357/abbc1b}, \href {https://ui.adsabs.harvard.edu/abs/2020ApJ...904...26Y} {904, 26}

\bibitem[\protect\citeauthoryear{{Yang} et~al.,}{{Yang} et~al.}{2022}]{yang2022}
{Yang} L.,  et~al., 2022, \mn@doi [\apj] {10.3847/1538-4357/ac7b2e}, \href {https://ui.adsabs.harvard.edu/abs/2022ApJ...935..121Y} {935, 121}

\bibitem[\protect\citeauthoryear{{Yokoyama} \& {Ohira}}{{Yokoyama} \& {Ohira}}{2023}]{yokoyama2023}
{Yokoyama} S.~L.,  {Ohira} Y.,  2023, \mn@doi [\mnras] {10.1093/mnras/stad1596}, \href {https://ui.adsabs.harvard.edu/abs/2023MNRAS.523.3671Y} {523, 3671}

\bibitem[\protect\citeauthoryear{{Zhou}, {Guo}, {Liu}, {Yue}, {Xu}  \& {Chen}}{{Zhou} et~al.}{2013}]{zhou2013}
{Zhou} J.,  {Guo} Q.,  {Liu} G.-C.,  {Yue} B.,  {Xu} Y.-D.,   {Chen} X.-L.,  2013, \mn@doi [Research in Astronomy and Astrophysics] {10.1088/1674-4527/13/4/001}, \href {https://ui.adsabs.harvard.edu/abs/2013RAA....13..373Z} {13, 373}

\bibitem[\protect\citeauthoryear{{Zhu} et~al.,}{{Zhu} et~al.}{2024}]{zhu2024}
{Zhu} Y.,  et~al., 2024, \mn@doi [\mnras] {10.1093/mnrasl/slae061}, \href {https://ui.adsabs.harvard.edu/abs/2024MNRAS.533L..49Z} {533, L49}

\bibitem[\protect\citeauthoryear{{{\v{D}}urov{\v{c}}{\'\i}kov{\'a}} et~al.,}{{{\v{D}}urov{\v{c}}{\'\i}kov{\'a}} et~al.}{2024}]{durovcicova2024}
{{\v{D}}urov{\v{c}}{\'\i}kov{\'a}} D.,  et~al., 2024, \mn@doi [\apj] {10.3847/1538-4357/ad4888}, \href {https://ui.adsabs.harvard.edu/abs/2024ApJ...969..162D} {969, 162}

\bibitem[\protect\citeauthoryear{{van der Walt}, {Colbert}  \& {Varoquaux}}{{van der Walt} et~al.}{2011}]{vander2011}
{van der Walt} S.,  {Colbert} S.~C.,   {Varoquaux} G.,  2011, \mn@doi [Computing in Science and Engineering] {10.1109/MCSE.2011.37}, \href {https://ui.adsabs.harvard.edu/abs/2011CSE....13b..22V} {13, 22}

\makeatother
\end{thebibliography}





\bsp	
\label{lastpage}
\end{document}